\documentclass[twocolumn]{aastex631}
\usepackage[utf8]{inputenc}
\usepackage{comment}
\usepackage{autobreak}
\usepackage{amsmath}
\usepackage{color}
\usepackage{threeparttable}
\usepackage{bm}

\newcommand\ltsima{$\; \buildrel <\over\sim \;$}
\newcommand\simlt{\lower.5ex\hbox{\ltsima}}
\newcommand\gtsima{$\; \buildrel >\over\sim \;$}
\newcommand\simgt{\lower.5ex\hbox{\gtsima}}

\shorttitle{Planet yields by the PRIME microlenisng survey}
\shortauthors{Kondo et al.}

\begin{document}

\title{Prediction of Planet Yields by the PRime-focus Infrared Microlensing Experiment Microlensing Survey}

\email{ikondoh@iral.ess.sci.osaka-u.ac.jp}

\author{Iona Kondo}
\affiliation{Department of Earth and Space Science, Graduate School of Science, Osaka University, Toyonaka, Osaka 560-0043, Japan}

\author{Takahiro Sumi}
\affiliation{Department of Earth and Space Science, Graduate School of Science, Osaka University, Toyonaka, Osaka 560-0043, Japan}

\author{Naoki Koshimoto}
\affiliation{Code 667, NASA Goddard Space Flight Center, Greenbelt, MD 20771, USA}
\affiliation{Department of Astronomy, University of Maryland, College Park, MD 20742, USA}

\author{Nicholas J. Rattenbury}
\affiliation{Department of Physics, University of Auckland, Private Bag 92019, Auckland, New Zealand}

\author{Daisuke Suzuki}
\affiliation{Department of Earth and Space Science, Graduate School of Science, Osaka University, Toyonaka, Osaka 560-0043, Japan}

\author{David P. Bennett}
\affiliation{Code 667, NASA Goddard Space Flight Center, Greenbelt, MD 20771, USA}
\affiliation{Department of Astronomy, University of Maryland, College Park, MD 20742, USA}

\begin{abstract}

The PRime-focus Infrared Microlensing Experiment (PRIME) will be the first to conduct a dedicated near infrared (NIR) microlensing survey by using a 1.8m telescope with a wide field of view of 1.45 ${\rm deg^{2}}$ at the South African Astronomical Observatory (SAAO).
The major goals of the PRIME microlensing survey are to measure the microlensing event rate in the inner Galactic bulge to help design the observing strategy for the exoplanet microlensing survey by the {\it Nancy Grace Roman Space Telescope} and to make a first statistical measurement of exoplanet demographics in the central bulge fields where optical observations are very difficult owing to the high extinction in these fields.
Here we conduct a simulation of the PRIME microlensing survey to estimate its planet yields and determine the optimal survey strategy, using a Galactic model optimized for the inner Galactic bulge. 
In order to maximize the number of planet detections and the range of planet mass, we compare the planet yields among four observation strategies. 
Assuming {the \citet{2012Natur.481..167C} mass function as modified by \citet{2019ApJS..241....3P}},
we predict that PRIME will detect planetary signals for $42-52$ planets ($1-2$ planets with $M_p \leq 1 M_\oplus$,  $22-25$ planets with mass $1 M_\oplus < M_p \leq 100 M_\oplus$, $19-25$ planets  $100 M_\oplus < M_p \leq 10000 M_\oplus$), per year depending on the chosen observation strategy.

\end{abstract}


\section{Introduction} \label{sec:intro}

The number of the detection of exoplanets has exceeded 5,000.  Most of these have been discovered via transit and radial velocity methods and  have orbital radii and masses different from those of the solar system planets.
The microlensing method, in contrast, is complementary to the other methods because it is sensitive to Earth-mass planets \citep{1996ApJ...472..660B} beyond the snow-line \citep{1992ApJ...396..104G}, as well as to free floating planets that are not orbiting a host star \citep{2011Natur.473..349S,2017Natur.548..183M,2022arXiv220403269G}. The snow-line represents the boundary {in the protoplanetary disk} where ${\rm H_2O}$ becomes ice, outside of which planet formation is predicted to be most active according to the core accretion model \citep{1993prpl.conf.1061L,1996Icar..124...62P}.
Currently, there are three optical microlensing survey projects; the Microlensing Observations in Astrophysics (MOA; \citealp{2001MNRAS.327..868B,2003ApJ...591..204S}), the Optical Gravitational Lensing Experiment (OGLE; \citealp{2015AcA....65....1U}) and the Korea Microlensing Telescope Network (KMTNet; \citealp{2016JKAS...49...37K}). 
Thanks to these survey observations and other follow-up observations, the total number of planets detected via microlensing is 141 as of 2022 November 2\footnote{https://exoplanetarchive.ipac.caltech.edu/docs/counts\_detail.html}. Statistical analyses using microlensing planets provide important findings such as cold planet frequency \citep{2016ApJ...833..145S} and constraints on the dependence of cold planet frequency on the Galactic location \citep{2021ApJ...918L...8K}. 
\citet{2016ApJ...833..145S} measured the mass-ratio function of planets beyond the snow-line using 29 planets discovered by the MOA and other optical microlensing surveys. 
{They found a break, and likely peak in the mass-ratio function near a Neptune mass for the first time.}
However, there is still a large degree of {uncertainty in the location of the break (or peak) in the planet mass-ratio distribution} owing to the lack of low-mass planets in their analysis. Recently \citet{2022MNRAS.515..928Z} have suggested 
a possibility that low-mass planets are more abundant than previous results. {Their analysis used 13 planets including small mass-ratio planets detected by KMTNet, but did not correct for detection efficiencies.}
\citet{2021ApJ...918L...8K} used the statistical samples in \citet{2016ApJ...833..145S} and showed that there is no strong dependence of the cold planet frequency on the Galactocentric distance.

The inner bulge ($|b| \simlt 2^\circ$) regions including the Galactic center have remained hidden for the current microlensing survey owing to high extinction. However, these regions are interesting because this is where we expect to find microlensing events in large quantities because of the high stellar density \citep{1995ApJ...446L..71G}. In the near infrared (NIR), light can penetrate through the dust in this region. Comparing the measurements of the planet frequency using an NIR microlensing survey with that determined by the present optical survey, the dependency of planet occurrence on the Galactic structure can be measured, which provides key insights into planetary formation and its history in the Galaxy.

So far, hundreds of microlensing events were discovered in the inner bulge region by the two NIR surveys, VISTA Variables in the Via Lactea Survey (VVV; \citealp{2010NewA...15..433M}) and the United Kingdom Infrared Telescope (UKIRT) Microlensing Survey \citep{2017AJ....153...61S, 2018ApJ...857L...8S}. The VVV survey {conducted} an NIR survey toward the inner Galactic bulge including the Galactic central region and adjacent region of the Galactic plane by using the Visible and Infrared Survey Telescope for Astronomy (VISTA), a 4 m telescope with the 1.6 deg$^2$ field of view (FOV) VISTA InfraRed Camera (VIRCAM; \citealp{2010Msngr.139....2E}) at ESO's Cerro Paranal Observatory in Chile. Although there are multiple epochs in $K_S$-band, the survey is not designed for microlensing and the observation cadence was irregular ($ 1/{\rm day}$ at best), {which is generally inadequate to detect microlensing light curves with features due to planets.} However, their survey is sufficient to reveal the number of microlensing events as a function of Galactic longitude and Galactic latitude. They found the Galactic longitude distribution $(-10.0^{\circ} < l < 10.44^{\circ})$ by using 630 microlensing events discovered during $2010-2015$ \citep{2018ApJ...865L...5N} and the Galactic latitude distribution $(-3.7^{\circ} < b < 3.9^{\circ})$ using 360 microlensing events \citep{2020ApJ...889...56N}.
From 2015 to 2018, the UKIRT Microlensing Survey \citep{2017AJ....153...61S} conducted a microlensing exoplanet survey toward the inner Galactic bulge by using the UKIRT 3.8 m telescope on Mauna Kea, Hawaii with a 0.8 deg$^2$ FOV infrared camera, Wide Field Camera (WFCAM). The UKIRT microlensing survey observed in $H$- and $K$-band filters. UKIRT-2017-BLG-001Lb \citep{2018ApJ...857L...8S} is the first planet that was found near the Galactic center at $(l,b)$ $=$ $(-0.12^{\circ},-0.33^{\circ})$ with a high extinction of $A_K =1.68$. The discovery of UKIRT-2017-BLG-001Lb demonstrated that an NIR survey enables the detection of planets close to the Galactic center with high extinction. {Although the above observations have been made, there are still no measurements of microlensing event rates and planet frequency in the inner Galactic bulge.}

The {\it Nancy Grace Roman Space Telescope} is NASA’s next flagship mission \citep{2015arXiv150303757S}, which is planned to launch in late 2026. It will be placed in a halo orbit around the second Sun-Earth Lagrange Point (L2). The main uses of $Roman$ are to study dark energy and to conduct a statistical census of exoplanets by conducting a microlensing survey. $Roman$ comprises a 2.4 m telescope with a 0.281 deg$^2$ wide FOV camera. The $Roman$ Galactic Exoplanet Survey \citep{2002ApJ...574..985B,2010arXiv1012.4486B} will comprise 15 minute cadence observations over a few square degrees toward the inner Galactic bulge with a wide W149 filter ($1-2$ $\mu$m). Thanks to the high photometric accuracy and continuous observations during {$\sim 72$} days in each of six seasons over five years, $Roman$ will detect $\sim 1400$ cold exoplanets with masses greater than that of Mars ($\sim 0.1 M_\oplus$) including 300 planets with mass of less than $3M_\oplus$ \citep{2019ApJS..241....3P}. In addition, \citet{2020AJ....160..123J} shows that $Roman$ would detect $\sim 250$ free floating planets.

Prior to the microlensing survey by $Roman$, the PRime-focus Infrared Microlensing Experiment (PRIME) will start its NIR microlensing survey toward the inner Galactic bulge in 2023. PRIME will conduct a high-cadence wide FOV survey by using a 1.8m telescope (f/2.29) with 1.45 deg$^2$ (0.5''/pix) FOV  at Sutherland Observatory operated by the South African Astronomical Observatory (SAAO).
Half of the observation time will be used for the microlenisng planet survey towards the inner Galactic bulge. The other half will be used for other sciences, such as the transit surveys for M-dwarfs and the transient search for counterparts of high-z gamma-ray bursts and gravitational-wave events.

Here we present results of our simulations that compare four observation strategies for the PRIME microlensing survey and predict the planet yields. In Section \ref{sec-prime}, we introduce the PRIME microlensing survey. Then we explain the methodology of our simulations in order to calculate the detection efficiency of microlensing events and planets in Section \ref{sec-sim}. Next, we calculate star counts, microlensing event rate, detection efficiencies, {and detection number of microlensing events and planets for each line of sight over the inner Galactic bulge in Section \ref{sec-sta}. We present microlensing and planet yields depending on four observation strategies in Section \ref{sec-str}}. Finally, we discuss our results and summarize our conclusions in Section \ref{sec-dis} and Section \ref{sec-sum}.

\section{PRime-focus Infrared Microlensing Experiment (PRIME)}\label{sec-prime}
\subsection{The PRIME Microlensing Survey}

PRIME will be the first dedicated NIR microlensing experiment for the inner Galactic bulge. PRIME will use a NIR camera called PRIME-Cam, consisting of four Teledyne HgCdTe 4K x 4K photodiode array (H4RG-10) detectors with 10-micron pixels.
The primary passband for {the} microlensing survey is $H$-band and $Z$-, $Y$-, $J$-band filters are also used for color measurements. The current plan, which is assumed in our simulations, is that each observation epoch will be composed of twelve 9-second co-added dithered exposures and take $160$ sec including overheads (readout time per exposure, $3$ sec, slew time for dithering, $1$ sec, and slew time for the next field, $4$ sec) per exposure.
Parameters for the PRIME telescope and PRIME-Cam are summarized in Table \ref{Table_prime}. We note that some parameters in Table \ref{Table_prime} are current assumptions and are subject to change.

\subsection{The Goal of the PRIME Microlensing Survey}
The main goals of the PRIME microlensing survey are to measure the microlensing event rate in the inner Galactic bulge to help design the observing strategy for $Roman$'s exoplanet microlensing survey and to make a first statistical measurement of exoplanet demographics in the central bulge fields where optical observations are very difficult owing to the high extinction in these fields. By comparing with the planet frequency measured by visible observation, PRIME will reveal the Galactic distribution of planet frequency.
PRIME also helps to provide insight into the performance of the H4RG-10 detectors that $Roman$ will use. Moreover, after the $Roman$ telescope begins to observe, the simultaneous observations of PRIME and $Roman$ enable us to measure the microlensing parallax which gives us the mass and distance of lens systems. {In particular, observations where the baseline between the Earth and L2 is $\sim 0.01$ au have a sensitivity to the parallax measurements in the timing of a caustic crossing \citep{2020A&A...633A..98W}, which is just as sharp a feature as planetary signals, and the parallax measurements down to the free-floating planets regime \citep{2022A&A...664A.136B}.}

\begin{table}
\centering
\caption{Adopted Parameters of PRIME microlensing survey}
\label{Table_prime}
\begin{center}
\begin{threeparttable}
\centering
 \begin{tabular}{lc}
   \hline \hline
   Mirror diameter(m) & 1.8 \\
   Field of View (deg$^2$) & 1.45 \\
   Detectors & 4 $\times$ H4RG-10 \\
   Pixel Scale ($''/{\rm pixel}$) & 0.5\\
   Plate Scale ($\mu$m/pixel) & 10\\
   Primary bandpass ($\mu$m) & 1.64$\pm$0.30 ($H$-band)\\
   \hline
   Exposure time (s) & 9 \\ 
   Readout number & 3 \\
   Stack number & 12 \\
   Readout noise(counts/pixel)\tnote{$^a$}  & 12.12\\
   Dark(counts/pixel/s)\tnote{$^a$}  & 0.130 \\
   \hline
   QE  & 0.88\tnote{$^b$} \\
   Throughput, $\eta$ & 0.78\tnote{$^c$}\\
   Thermal background (counts/pixel/s)  & 500\tnote{$^d$} \\
   Sky background (counts/pixel/s)  & {3400--9400}\tnote{$^e$}\\
   Limiting magnitude (mag) & 18.5\tnote{$^f$} ($H$-band)\\
   Saturation limit (mag) & 11.0 ($H$-band)\\
  \hline \hline
 \end{tabular}
\begin{tablenotes}
  \item[$^a$] The readout noise and dark count value is assumed to be the same as those of the $Roman$ telescope as shown in \citet{2019ApJS..241....3P}.
  \item[$b$] Assumed QE in $H$-band.
  \item[$c$] Throughput is estimated by multiplying the assumed transmittance of the atmosphere, the measured reflectance of the primary mirror, the measured transmittance of AR coatings, the measured transmittance of the filters, and the assumed detector QE. 
  \item[$d$] Assumed thermal background at 290 K in $H$-band. 
  \item[$e$] Assumed sky background in $H$-band, {corresponding to $13.0-14.2$ ${\rm mag/arcsec}^2$. These limits are derived based on the sky emission from Cerro Pachon at the Gemini Observatory (https://www.gemini.edu).}
  \item[$f$] Faint magnitude limit for a $5\sigma$.
\end{tablenotes}
\end{threeparttable}
\end{center}
\end{table}

\section{Simulations} \label{sec-sim} 
Although an expected microlensing event rate of each field in the inner bulge can be calculated by a model of our Galaxy, we need a survey simulation to obtain detection efficiencies of (i) microlensing events and (ii) planetary events to calculate how many microlensing events and planets are expected to be found by PRIME.

In this section, we present procedure of a Monte Carlo simulation for one year of PRIME observations toward the inner Galactic bulge with 16, 32, 48, and 96 minute cadence observations to estimate the detection efficiencies as a function of field coordinate and observation cadence.

\subsection{Simulation overview}

\begin{figure*}[th]
      \hspace*{-1cm}
      \includegraphics[scale=0.50]{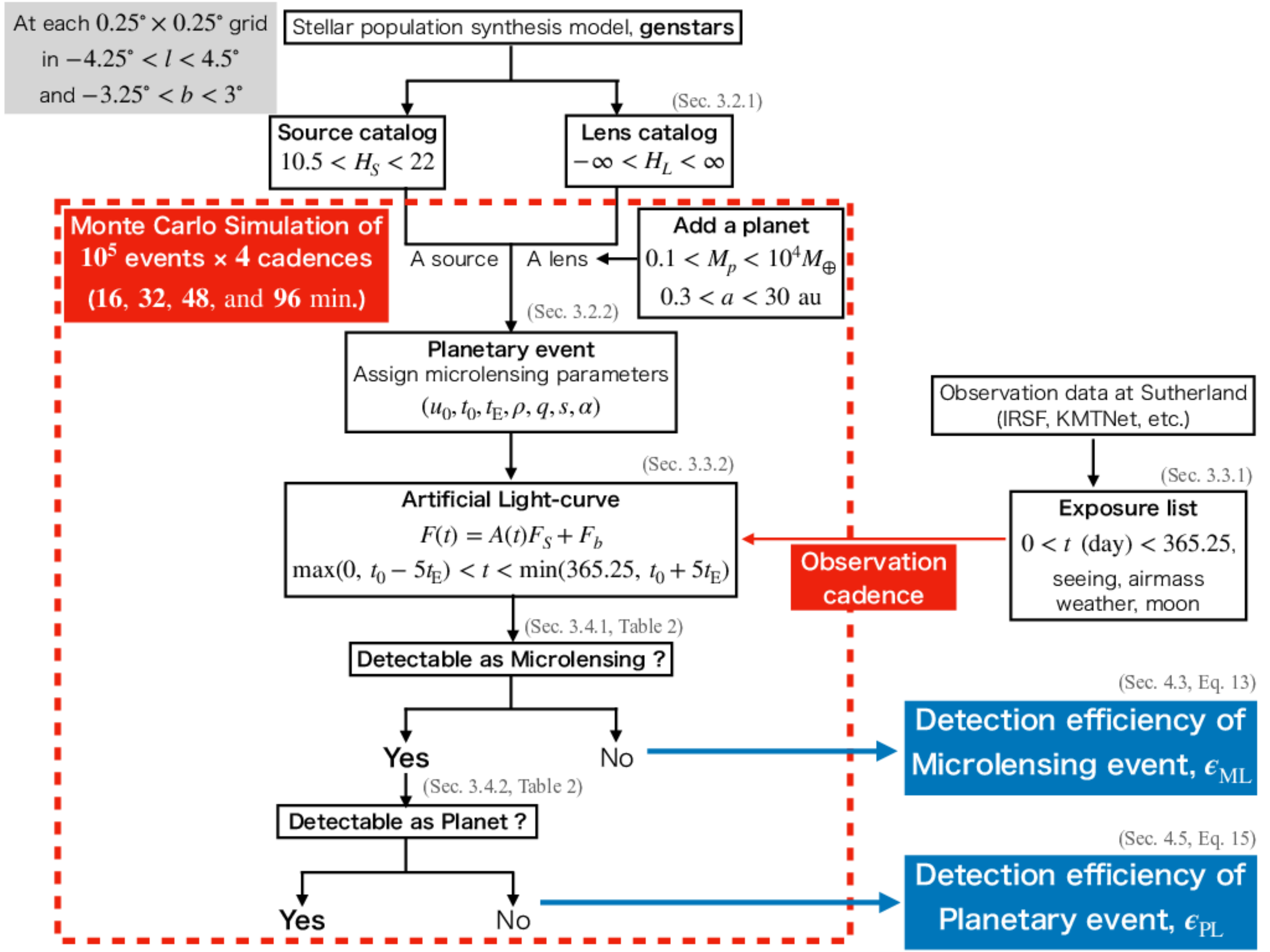}
  \caption{Schematic view of our simulation to estimate the detection efficiency of both microlensing events and planets at specific $(l,b)$. For each Galactic coordinate and for each observation cadence, a Monte Carlo simulation is performed to calculate the detectability of one hundred thousand microlensing events.}
  \label{fig_sche}
\end{figure*}

Figure \ref{fig_sche} shows a schematic view of our simulation. For each Galactic coordinate and for each observation cadence, a Monte Carlo simulation is performed to calculate the detectability of one hundred thousand microlensing events. A brief explanation of each procedure is presented in the following.

First, we randomly select source and lens objects from each star catalog at specific Galactic coordinates, $(l,b)$, generated from a stellar population synthesis model in our Galaxy. We then assign parameters for single-lens microlensing and binary-lens microlensing with planetary mass-ratios. Synthetic light curves are generated. Each light curve is then modified according to the observation cadence, the parameters of PRIME-Cam and telescope, and observation conditions at Sutherland. Finally, based on the detection criteria, we will examine whether the microlensing events and the planetary signatures can be detected.

\subsection{Simulation of planetary microlensing events}\label{sec-sim-ml}
In this section, we describe how to simulate planetary microlensing light-curves.
First, we generate a microlensing event by randomly drawing lens and source stars from catalogs of lens and source stars created by the Galactic model and adding a planet to the lens. Then we compute the parameters of single-lens and binary-lens models which are associated with the physical parameters assigned to the combination of the source and lens. Then, we calculate the magnification of that event as a function of time.

\subsubsection{Galactic model and Catalogs of source and lens} \label{sec-galmod}

Koshimoto et al. (in prep). developed a stellar population synthesis tool, \texttt{genstars}\footnote
{The software is available via Zenodo \citep{kos22} or \dataset[https://github.com/nkoshimoto/genstars]{https://github.com/nkoshimoto/genstars}.}, which uses a modified version of the Galactic model by \citet{2021ApJ...917...78K}. The modified model is applicable for the inner bulge region because it has a nuclear stellar disk (NSD) structure based on the NSD model by \citet{sor22}. The NSD is not included in other population synthesis tools such as the Besan\c{c}on model \citep{rob03, 2012A&A...538A.106R} or \texttt{Galaxia} \citep{sha11}. Thus, \texttt{genstars} is currently only the public population synthesis tool suitable for our simulation toward the inner bulge region.

Note that we use a slightly different version of \texttt{genstars} from the public version, where the center of our Galaxy is at $(l, b) = (0, 0)$ rather than at Sgr A* at $(l, b) = (-0.056^\circ, -0.046^\circ)$ \citep{rei04}. The central shift slightly affects our simulation results in the inner NSD region or central $\sim$ 0.5 deg$^2$. However, the influence is negligible compared to other issues such as the underestimation of extinction in the Galactic central region which is shown in Koshimoto et al. (in prep). This version of their Galactic model  will hereafter be referred to as KGM.

In order to simulate the combination of a source and a lens for microlensing events, we use two star catalogs. The first list, the list of {sources}, is selected by specifying a range of magnitudes, $10.5 < H_S < 22$ {in the Vega magnitude system} within 16 kpc from the Sun. The source list includes stars fainter than PRIME's limiting magnitude, $H_{\rm lim}\sim 18.5$, because they can become bright enough to be detected if sufficiently magnified. The second list, the list of lenses, is selected without magnitude limitations ($-\infty < H_L < \infty$), i.e., including dark objects such as brown dwarfs, white dwarfs, neutron stars, and black holes. 
{Each list contains the following physical parameters of sources or lenses: the magnitude, mass, radius, distance, and proper motions.}

\subsubsection{Microlensing parameters} \label{sec-simuparams}

A microlensing event occurs when a foreground lens star passes close to the line of sight between an observer and a background source star. The gravity of the lens star bends the light from the source star and magnifies its brightness. The angular Einstein ring radius is given by,
\begin{equation}\label{eq-thetae}
    \theta_{\rm E} = \sqrt{\kappa M_L\pi_{\rm rel}},
\end{equation}
where $M_L$ is the mass of the lens object, and $\kappa = 4G(c^2{\rm \ au})^{-1} = 8.14 {\rm \ mas}M_{\odot}^{-1}$. When the distance from the observer to the lens and source are represented by $D_L$ and $D_S$, respectively, the lens-source relative parallax is $\pi_{\rm rel}=1{\rm \ au}(D_L^{-1} - D_S^{-1})$.

The magnification of the single-lens light-curve model depends on three parameters: the time of lens-source closest approach $t_{0}$, the impact parameter in units of the Einstein radius $u_{0}$, and the Einstein radius crossing time $t_{\rm E}$. 
We also include the finite source effects and introduce one parameter: the ratio of the angular source size to the angular Einstein radius, $\rho$.

We assume uniform distributions of $t_0$ and $u_0$:
\begin{align}
   0\leq t_0\leq T_{\rm  obs}, \\
    0\leq u_0 \leq u_{0,\rm max},
\end{align}
where we adopt the survey duration $T_{\rm obs} = 365.25$ day. We also adopt the maximum value of impact parameter $u_{0,\rm max}=1.0$. The events with $u_{0,\rm max}>1.0$ do not significantly affect the final result because the detection efficiency is lower owing to the low magnification. $t_{\rm E}$ and $\rho$ are derived from the physical parameters assigned to the combination of the source and lens,

\begin{align}
   t_{\rm E} = \frac{\theta_{\rm E}}{\mu_{\rm rel}}\label{eq-te}\\
    \rho = \frac{\theta_*}{\theta_{\rm E}},
\end{align}
where $\mu_{\rm rel}$ is the lens-source relative proper motion drawn from the velocity distribution in the Galactic model. The angular radius of the source star $\theta_* = R_*/D_S$, where $R_*$ is the radius of the source star estimated from the source magnitude from \texttt{genstars}.
Note that the microlensing event rate is not equal among all the source--lens pairs picked up from the catalogs because it is $\propto \mu_{\rm rel} \theta_{\rm E}$. We will later add this weight when considering the statistics of simulated events.

The magnification of the binary-lens model requires three additional parameters; the planet-host mass ratio, $q$, the planet-host separation in units of the Einstein radius, $s$, the angle between the trajectory of the source and the planet-host axis, $\alpha$.
The mass ratio and the planet-host separation are given by
\begin{align}
   q &= \frac{M_p}{M_h} \\
     s &= \frac{a_{\perp}}{D_L \theta_{\rm E}},
\end{align}
where $M_p$ and $M_h$ are the mass of the planet and the host star, respectively. Assuming a circular orbit, the projected orbital separation $a_{\perp}=a\sqrt{1-\cos^2{\zeta}}$, where $a$ is semi major axis and $\zeta$ is the angle between the plane of the sky and the binary-axis at a given time. We use a uniform distribution of $\cos{\zeta}$ assuming a circular planetary orbit that is inclined randomly to the line of sight.
We use 21 fixed values of planetary mass distributed logarithmically in the range $0.1<M_p<10^4 M_\oplus$ (0.10, 0.18, 0.32, ... , 10000 $M_\oplus$) and 15 fixed values of semi major axis in the range $0.3<a<30$ au (0.3, 0.42, 0.58, ... , 30 au). We also assume a uniform distribution of $0<\alpha<360$.

\subsubsection{Magnification calculation} \label{sec-simumag}
We calculate the magnification of the single-lens model as a function of time, using either the \citet{2004ApJ...603..139Y} or the \citet{2009ApJ...695..200L} method depending on the value of $\rho$ for the calculation of the finite source with limb darkening as implemented in \texttt{MulensModel} \citep{2019A&C....26...35P}.
In order to calculate the magnification of the binary-lens model, we use the advanced contour integration method as implemented in \texttt{VBBinaryLensing} \citep{2010MNRAS.408.2188B, 2018MNRAS.479.5157B}. In our simulations, we do not consider higher-order effects such as parallax, xallarap, or lens orbital motion.

We note that the magnification of the binary-lens model are calculated to generate synthetic data points in Section \ref{sec-gen-data} and to examine the validity of planetary signatures in Section \ref{sec-det-cri-ml}. The magnification of the single-lens model are calculated to investigate the detectability of microlensing events and planetary signatures by the $\chi^2$ value of the single-lens
model in Section \ref{sec-det-cri-ml}.

\subsection{Generate synthetic data points}\label{sec-gen-data}
After generating the microlensing models, the next step is to model how the microlensing events are observed by PRIME.
We generate the synthetic data points with 16, 32, 48, and 96 minute cadences.

\subsubsection{Exposure list}

First of all, we make an exposure list of observational parameters such as seeing and airmass for each exposure time ($\sim 160$ sec). In order to reproduce actual observations, we consider the visibility of the Galactic center, weather, and the days of full moon at Sutherland.
The observation toward the inner Galactic bulge is assumed to be conducted when the Sun's altitude is more than 12 degrees below the horizon and when the Galactic center's altitude is more than 20 degrees. Then, we remove the days of the bad weather and three days across the full moon from the set of observable times, based on observation statistics\footnote{https://kmtnet.kasi.re.kr/kmtnet-eng/observing-statistics-of-three-sites/} and online data\footnote{https://kmtnet.kasi.re.kr/ulens/} over $2016-2018$. The simulated observable time accounts for $\sim 55-60$ \% of the whole night time of the bulge season. 

After making the exposure list of the epochs when the Galactic center is visible, we assign the value of airmass and seeing to each exposure time. We calculate airmass from the altitude of the Galactic Center, ${\rm airmass} =\sec(z)$, where $z$ is the zenith angle. We draw the seeing values from the log normal distribution presented in \citet{2007PASJ...59..615K}. That work provides an observational seeing distribution under certain airmass conditions obtained observations of the Large Magellanic Cloud from Sutherland with the InfraRed Survey Facility (IRSF). We also consider the airmass dependence of the seeing, ${\rm airmass}^{0.6}$, given by \citet{1982ARA&A..20..367W}.

\subsubsection{Flux determination}
Now we have the exposure list, where the observational parameters such as exposure epoch, seeing, and airmass are assigned. Then we calculate the flux for each observable data point of a microlensing event. The PRIME photometry will be reduced by using an implementation of the MOA Difference Imaging Analysis (DIA) pipeline \citep{2001MNRAS.327..868B}. Since the microlensing survey is conducted toward the inner Galactic bulge, where {the surface density of stars is expected to be high}, aperture photometry and point-spread function fitting photometry are known to be less effective in these crowded fields.
With the magnification of the source flux as a function of time, $A(t, {\bm x})$, which is defined the microlensing parameters, ${\bm x} = (u_0, t_0,t_{\rm E},\rho,q,s,\alpha)$ described in Section \ref{sec-simuparams}, the total flux of the magnified source, $F(t)$, is given by 
\begin{equation}
F(t)= A(t, {\bm x})F_s +F_b,
\end{equation}
where $F_s$ is the baseline flux of the source star, and $F_b$ is the blend flux which can, in principle include the lens flux. 

When we simulate data points for each microlensing event, data points are generated during $T_{\rm min} <t<T_{\rm max}$, where $T_{\rm min} = t_0-5 t_{\rm E}$ and $T_{\rm max} = t_0+5 t_{\rm E}$. If $T_{\rm min}<0$, we use $T_{\rm min}=0$ and if $T_{\rm max}>365.25$, we use $T_{\rm max}=365.25$.

We calculate the source flux, $F_s$, by combining the $H$-band magnitude of the source star, $H_S$ generated from \texttt{genstars} with the throughput, $\eta$ in Table \ref{Table_prime}. 

To estimate the blending flux $F_b$, we calculate the lens flux from the $H$-band magnitude of the lens star, $H_L$, and the total flux of stars brighter than the limiting magnitude within the PSF, $F_{\rm bright}$. We derive $F_{\rm bright}$ by using the $H$-band images taken by the VVV survey fourth data release (DR4) \citep{2010NewA...15..433M}. We evaluate $F_{\rm bright}$ by subtracting the smooth background flux from the total flux in the region within the typical $H$-band seeing disc at Sutherland ($\sim 1.4^{\prime \prime}$). Then, the blending flux, $F_b$, can be obtained by adding the lens flux and $F_{\rm bright}$ contaminated in the event.

\begin{figure}[t]
    \centering
    \hspace*{-0.8cm}
    \includegraphics[scale=0.24]{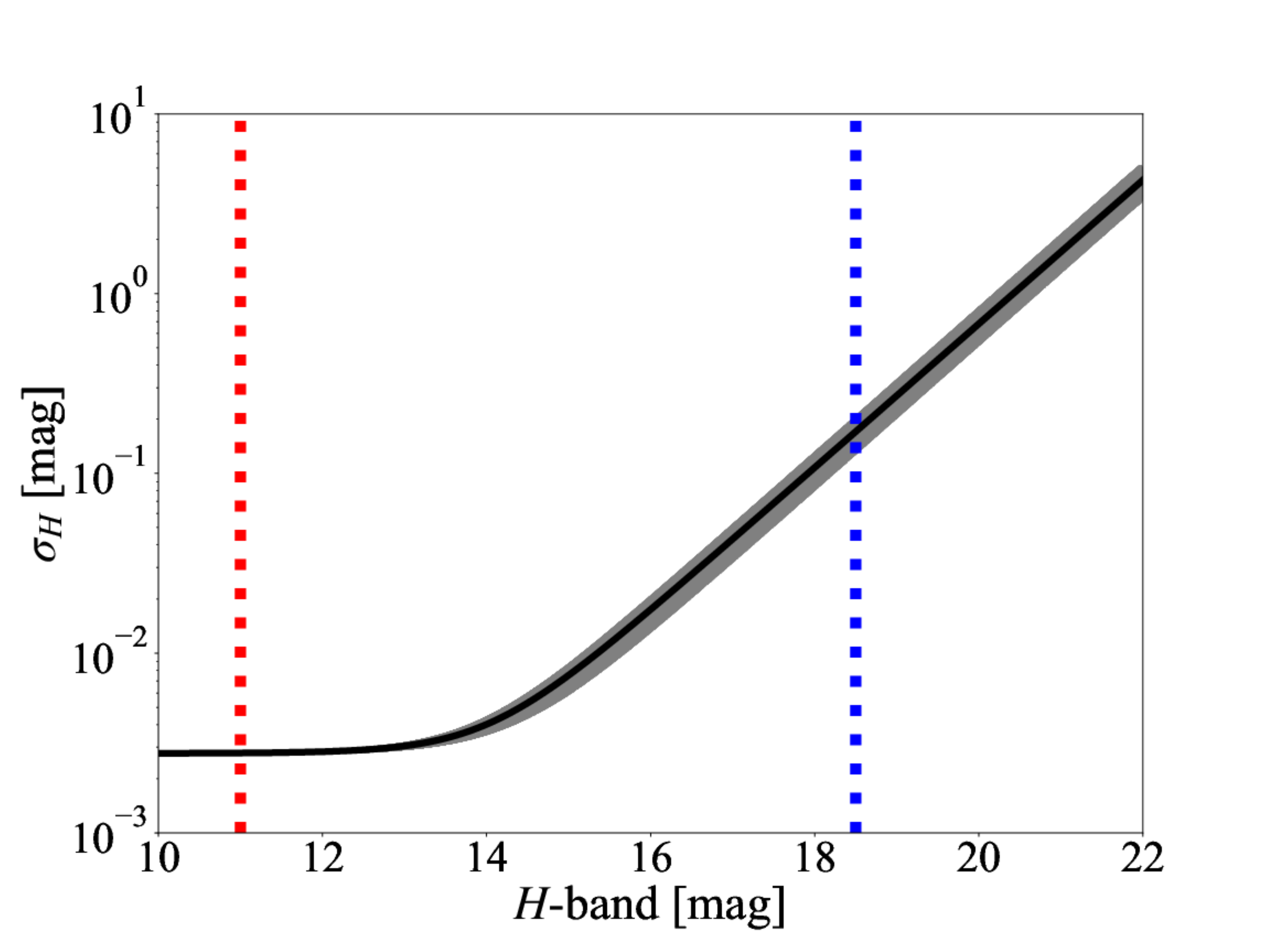}
  \caption{{Photometric precision of PRIME as a function of $H$-band magnitude when each observation epoch will be composed of twelve 9-second co-added dithered exposures and take 160 sec including overhead, assuming no blending and typical seeing. The black line shows the photometric precision assuming the sky background is 13.6 ${\rm mag/arcsec}^2$. The gray region shows the photometric precision assuming the sky background is $13.0-14.2$ ${\rm mag/arcsec}^2$. The red and blue dotted lines indicate the saturation limit in a single read and the faint magnitude limit for a $5\sigma$.}}\label{fig-photo}
\end{figure}

We evaluate the flux uncertainty $F_{\rm err}$ by quasi-smooth backgrounds such as sky backgrounds and faint unresolved stars, and instrumental backgrounds such as thermal background and dark current. These sources of error and their magnitudes are summarized in Table \ref{Table_prime}. 
{In ground-based observations, the brightness of the sky background is higher in the NIR wavelength than in the optical wavelength. In particular, intensities of the OH emission lines significantly dominate the sky background in the $H$-band. OH lines are known to fluctuate not only within the FOV but also throughout the night. In our simulation, we simulate those variations by randomly taking the sky brightness from a uniform range of $13.0-14.2$ ${\rm mag/arcsec}^2$ for each observation epoch because there is no measurement of the specific distribution of $H$-band sky brightness and its dependence on the observation conditions at Sutherland. 
We also consider that variations in the sky background due to changes in the moonlight are almost negligible because we exclude observations across a three day interval across the time of full moon. This is a conservative assumption because the moon's contribution to the sky background is minimal in the $H$-band when the separation angle between the target and the moon is more than 10 degrees \citep{2014NewA...28...63P}.}
{Although there may be systematic errors due to insufficient sky subtractions, DIA will deal with slight variations in sky background in actual observation. Thus, we do not take them into account in our simulations assuming that the sky background is successfully subtracted.}
The average flux of quasi-smooth background produced by faint unresolved stars, $F_{\rm faint}$ is estimated by the smooth background light in the region within the resolution of the simulation, $0.25^{\circ}\times 0.25^{\circ}$ using the $H$-band images in VVV DR4. 
We consider both Poisson noise from the total flux, $F(t)$, quasi-smooth backgrounds and instrumental backgrounds, and Gaussian noise from the readout noise.
It is known that the true photometric errors are underestimated owing to the crowded stellar fields, nearby bright stars, scintillation, and flat-fielding, etc. In order to include a fractional systematic uncertainty, we also add 0.3 \% of the magnitude in quadrature to each error. 



{The resultant photometric precision for each observation epoch as a function of $H$-band magnitude is shown in Figure \ref{fig-photo}, assuming no blending and typical seeing. Each observation epoch will be composed of twelve 9-second co-added dithered exposures and take 160 sec including overhead. As the gray area shows, photometric precision varies by up to 20\% with respect to the black line, depending on the value of sky brightness.}
{The typical photometric accuracies are $\sigma_{K_S}=0.01$ mag and $\sigma_{J,H} = 0.03$ mag for the VVV survey \citep{2020ApJ...893...65N}, and $\sigma_{Y,J,H}<0.02$ mag for the UKIRT Microlensing Survey \citep{2007MNRAS.379.1599L,2008MNRAS.391..136L}. The photometric precision of the PRIME microlensing survey is $\sigma_H<0.03$ mag for bright sources with $H<16.5$ mag. Moreover, the limiting magnitude of PRIME is $H_{\rm lim}\sim 18.5$ mag, which is brighter than limiting magnitudes\footnote{The limiting magnitudes are $H_{\rm lim}\sim 19.5$ mag for the VVV survey \citep{2019A&A...632A..85Z} and $H_{\rm lim}\sim 19.0$ mag for the UKIRT Microlensing Survey \citep{2007MNRAS.379.1599L}.} of those surveys. This is reasonable considering PRIME's smaller aperture than these two telescopes. Compared with those NIR surveys, PRIME has a comparable performance to those other NIR surveys, but will conduct the microlensing survey with much higher observation cadences, which is essential for the detection of planetary signals due to low-mass planets.}


\subsection{Detection Criteria}\label{sec-sim-de}

\begin{deluxetable*}{lll}
\tablecaption{Detection criteria \label{Table_criteria}}
\tablehead{
\colhead{level}  & \colhead{criteria} & \colhead{comments}  
}
\startdata
Microlensing   &  $\Delta\chi^2_{\rm ML}  > \Delta\chi^2_{\rm ML, th}=500$  & $\Delta \chi^2$ between the constant flux and single-lens models must be $> 500$ \\
               &  $N_{\rm data} >100$                                       &  Number of data points must be $> 100$\\
               &  $N_{\rm data, (t < t_0)} \geq 1$ \& $N_{\rm data, (t > t_0)} \geq 1$ & Data point(s) must exist before and after the peak time of the event\\
               & $A(t_{\rm max}) F_s / F_{\rm err} (t_{\rm max}) > 5$                             &  Maximum value of the source brightness at $t_{\rm max}$ must be $> 5$ times larger than \\
               &                                                            & the flux error at the time, $t_{\rm max}$  \\
               &    $N_{5\sigma} > 3$                                       & $> 3$ consecutive points with $> 5 \sigma$ deviation from the baseline must exist.\\
               \hline
  Planet       & $\Delta\chi^2_{\rm PL} > \Delta\chi^2_{\rm PL, th} = 160$  &  $\Delta \chi^2$ between the single-lens and binary-lens models must be $> 160$ 
\enddata
\end{deluxetable*}

\subsubsection{Microlensing event}\label{sec-det-cri-ml}
In order to detect planets via microlensing, it is required to detect both the microlensing event itself and to distinguish the planetary perturbations from the single-lens event.
We defined five criteria for the detection of microlensing events, which are summarized in Table \ref{Table_criteria}.
The first criterion is as follows,
\begin{equation}
    \Delta\chi^2_{\rm ML} \equiv \chi^2_{\rm const} - \chi^2_{\rm ML} > \Delta\chi^2_{\rm ML, th},
\end{equation}
where $\chi^2_{\rm const}$ and $\chi^2_{\rm ML}$ is the $\chi^2$ of the best-fit constant flux and best-fit single-lens model, respectively. We use $\Delta\chi^2_{\rm ML, th} = 500$.
The second criterion is that there must be more than 100 data points to guarantee modeling accuracy.
The third criterion is that there must be data points before and after the peak time of the event, which enhances the accuracy of the parameters measured from the light-curves.
The fourth criterion is that the maximum value of the source brightness must be $> 5$ times larger than the flux error at the time. {We note that this criteria is more conservative than criteria that is used in the analysis by KMTNet \citep{2022MNRAS.515..928Z,2022arXiv221012344Z}.}
The fifth criterion is that there are at least three consecutive points with the observed flux deviating from the constant baseline by more than $5 \sigma$. 
This requirement is intended to reduce the occasional artifacts on the baseline, like cosmic ray hit. Note that some events passed these criteria thanks to their planetary perturbation. Thus, even events with weak signals from the microlensing event itself have not been missed in our simulations if its planetary signature is sufficiently strong.

\subsubsection{Planetary Signature}\label{sec-det-cri-pl}
To estimate the expected yields of the planet detection by the PRIME microlensing survey, we need to set the planet detection criteria.
Our criterion for the detection of planetary signature is as follows,
\begin{equation}
    \Delta\chi^2_{\rm PL} \equiv \chi^2_{\rm ML} - \chi^2_{\rm PL} > \Delta\chi^2_{\rm PL, th},
\end{equation}
where $\chi^2_{\rm ML}$ and $\chi^2_{\rm PL}$ is the $\chi^2$ of the best-fit single-lens model and binary-lens model, respectively.
We use $\Delta\chi^2_{\rm PL, th} = 160$ following previous microlensing simulation studies (e.g., \citealp{2002ApJ...574..985B, 2013MNRAS.434....2P, 2014ApJ...794...52H}).

Although \citet{2016ApJ...833..145S} conducted their statistical analysis using a $\Delta\chi^2_{\rm PL}$ threshold of $100$ from only MOA survey data,
we use $\Delta\chi^2_{\rm PL, th}=160$ as a conservative assumption in order to consider uncertainties in our simulation.
We investigate the impact of changing $\Delta\chi^2_{\rm PL, th}$ on our simulation results. When we use $\Delta\chi^2_{\rm PL, th} = 100$, the detection efficiency of planetary signatures averaged over the planetary masses becomes $\sim12 \%$ higher than that of $\Delta\chi^2_{\rm PL, th} = 160$. 
As the result, the change of threshold slightly increases the planet detections described in Section \ref{sec-yield}.
We also estimate the detection efficiency of planetary signatures averaged over the planetary masses for $\Delta\chi^2_{\rm PL, th} = 300$ and find that the detection efficiency {is} $\sim 16\%$ lower than that of $\Delta\chi^2_{\rm PL, th} = 160$. Despite the lower detection rate, the number of Earth-mass planets to be detected is still more than one. Although the change of threshold affects the planet yields slightly, there is no significant change in the trend in the number of planet detections depending on observation strategies and our results in Section \ref{sec-yield}.

\begin{figure*}[h]
  \begin{minipage} [b]{0.45\linewidth}
    \centering
    \includegraphics[scale=0.18 ]{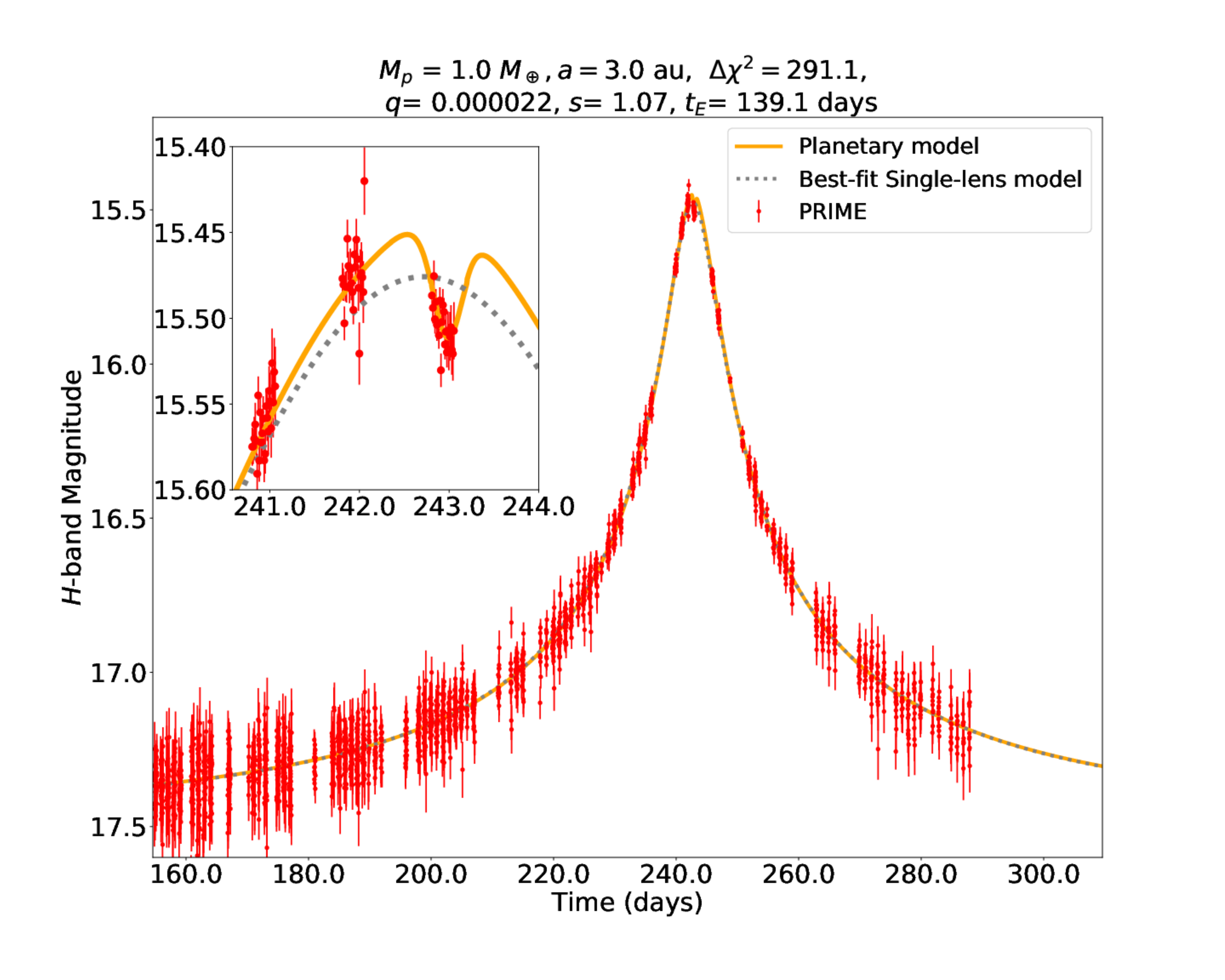}
  \end{minipage}
  \begin{minipage}[b]{0.45\linewidth}
    \centering
    \includegraphics[scale=0.18]{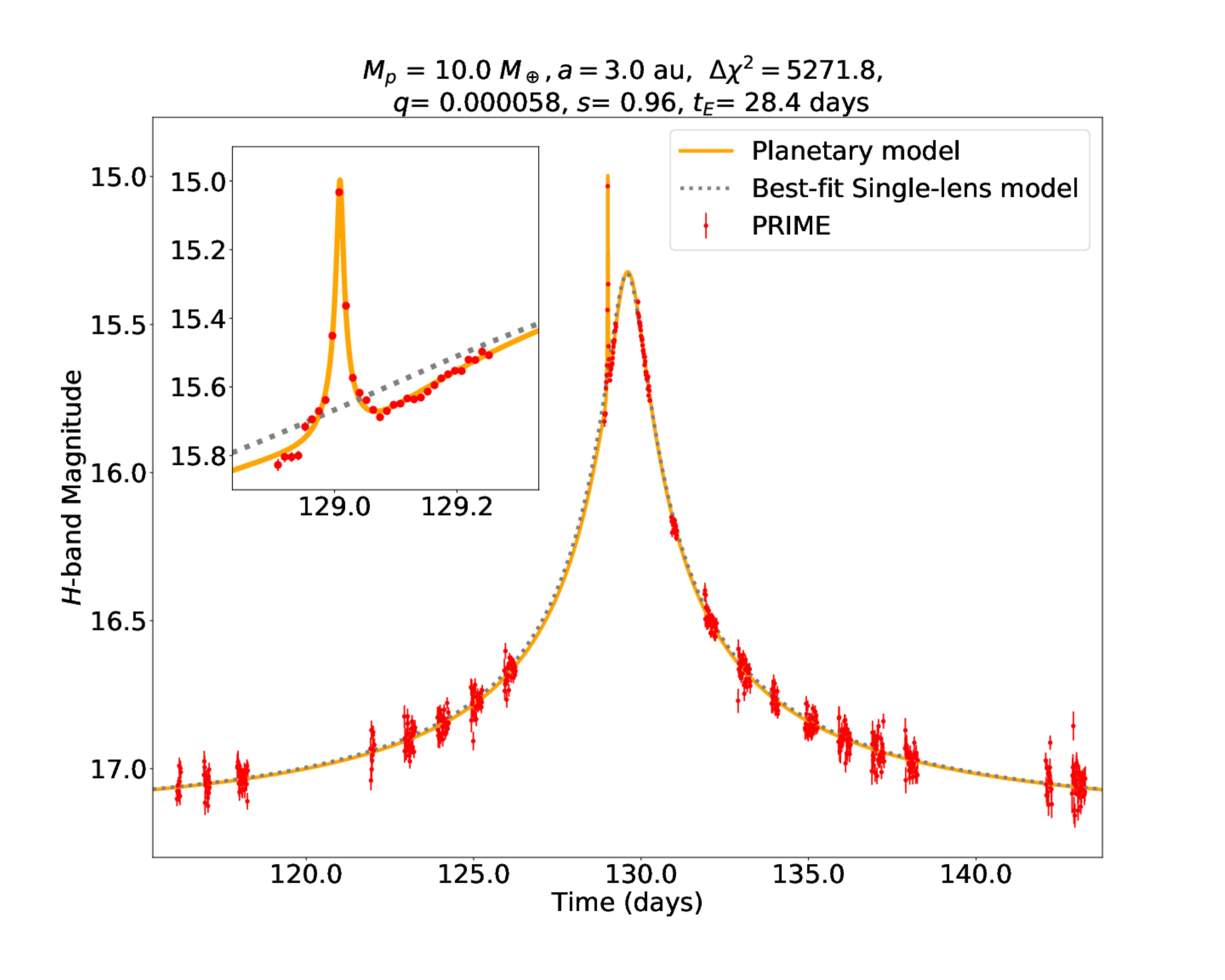}
  \end{minipage}\\
  
  \vspace{-0.8cm}
  
  \begin{minipage}[b]{0.45\linewidth}
    \centering
    \includegraphics[scale=0.18]{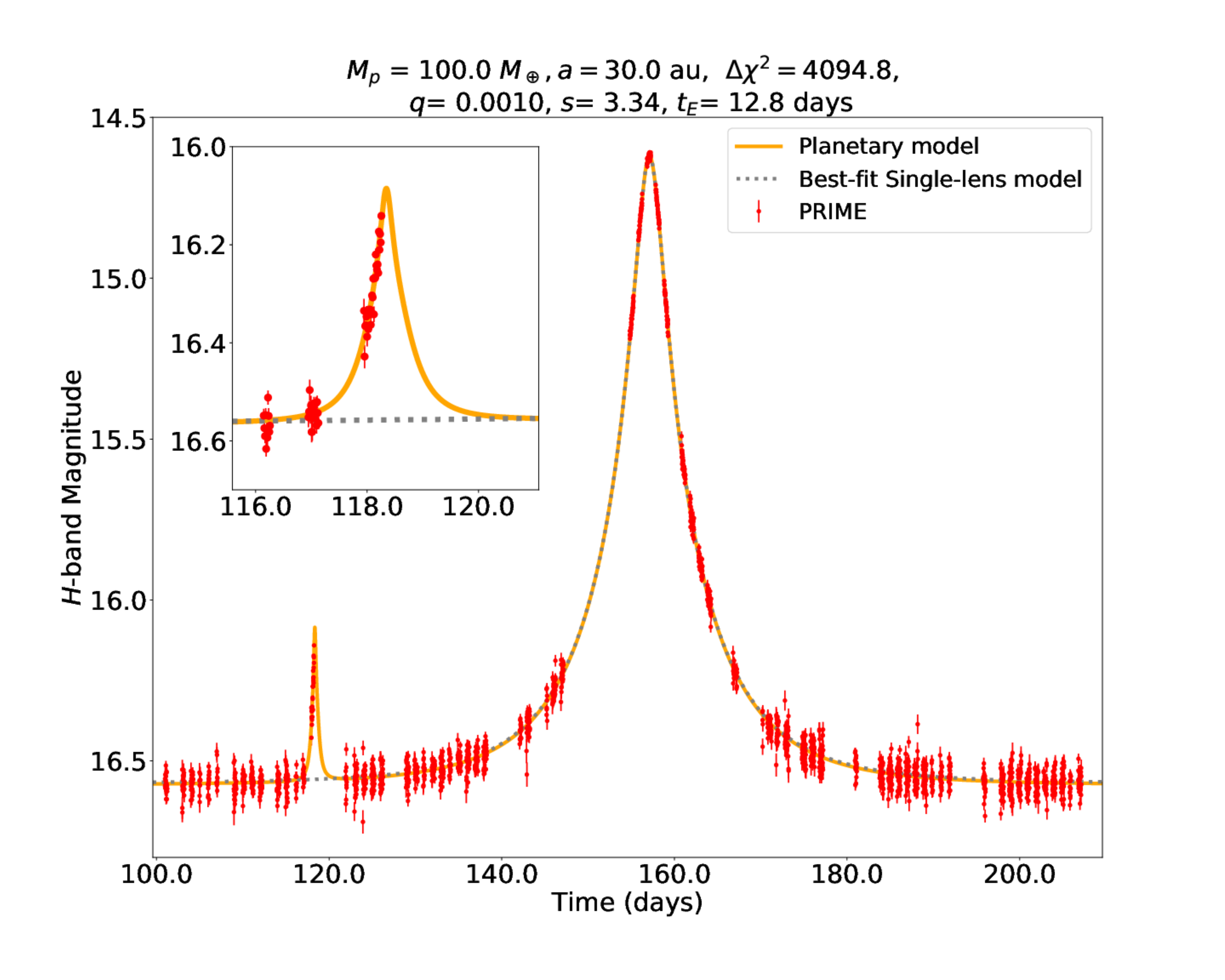}
  \end{minipage}
  \begin{minipage}[b]{0.45\linewidth}
    \centering
    \includegraphics[scale=0.18]{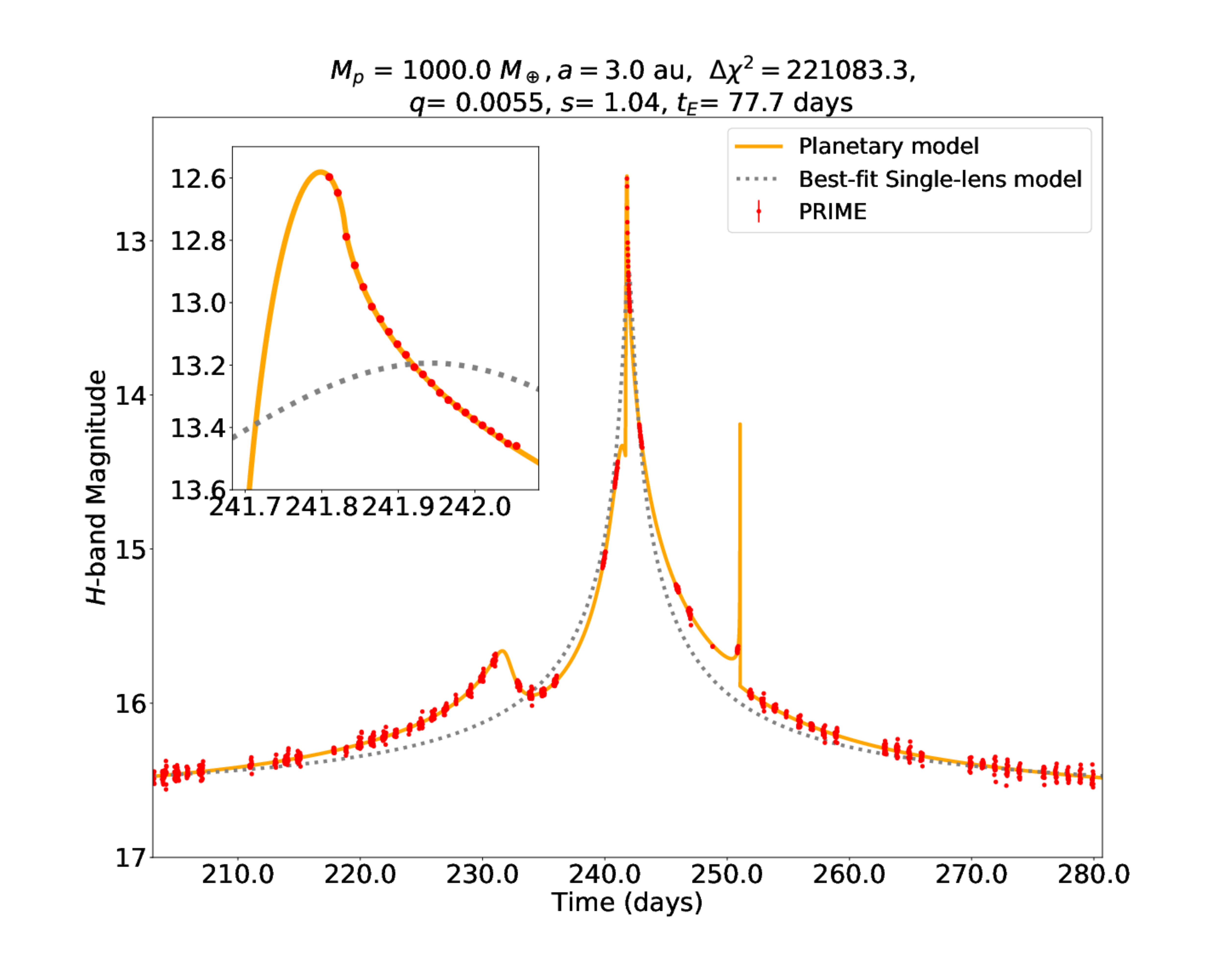}
  \end{minipage}\\

  \vspace{-0.8cm}
  \caption{Examples of simulated microlensing events whose planetary perturbation are detectable with the PRIME microlensing survey. The insets show the zoom-in of planetary signatures. The red dots show the synthetic data points with a 16 minute cadence. The planetary model for each event is shown in the orange line. The gray dotted lines show the best-fit single lens models.}\label{fig-lc-det}

\end{figure*}

\begin{figure}[ht]
    \centering
    \includegraphics[scale=0.18]{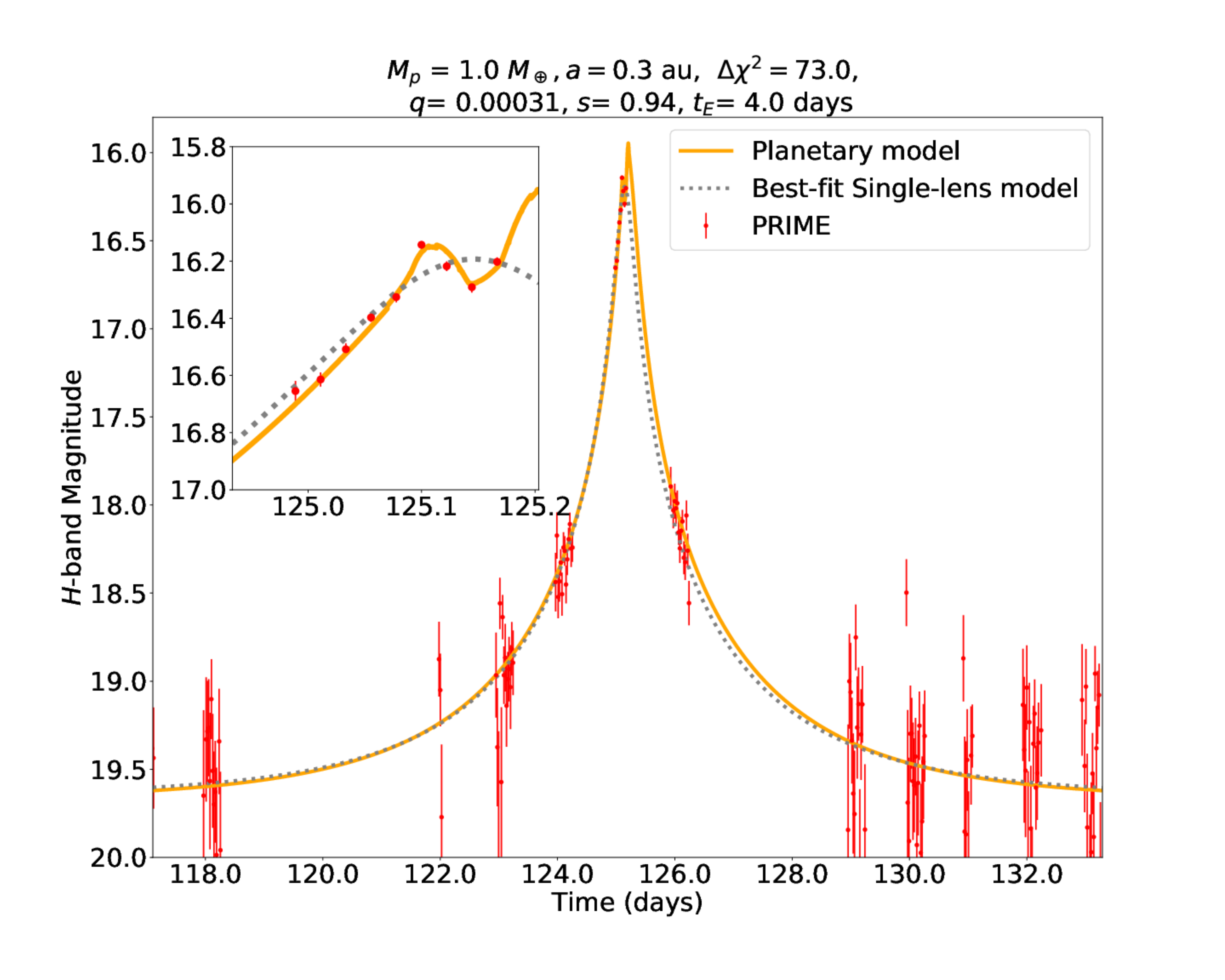}
    \\
    \centering
    \includegraphics[scale=0.18 ]{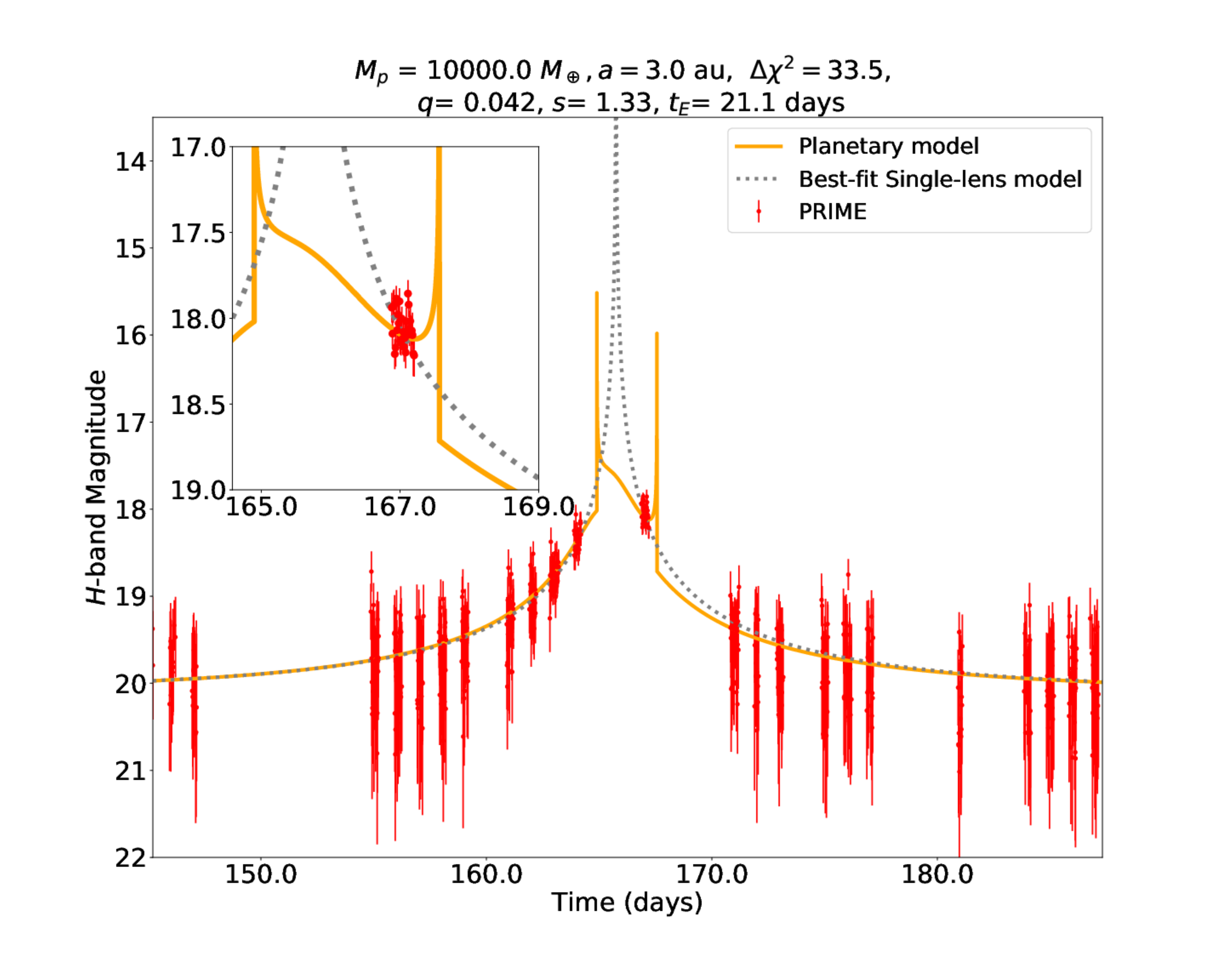}
  \caption{Same as Figure \ref{fig-lc-det}, but for the planetary microlensing events that do not pass the detection criteria of the planetary signatures. Observation cadence is 32 minutes in these examples.}\label{fig-lc-non}
\end{figure}

\subsection{Simulated light-curves}

Figure \ref{fig-lc-det} shows examples of simulated microlensing events in which the planetary signature can be detectable by PRIME. Although the duration of the significant deviation due to the low mass planet is only a few hours (top panels in Figure \ref{fig-lc-det}), the planetary signature is detectable if there are sufficient observation data. The detection efficiency for high mass planets is high because {the duration of the planetary perturbaiton is typically a few days} (bottom panels in Figure \ref{fig-lc-det}).

On the contrary, Figure \ref{fig-lc-non} shows examples of planetary events whose planetary signatures are missed in our simulation.
The artificial event in the top panel is located in a field observed with a 32 minute cadence. The duration of the signature due to a planet with mass of 1 $M_\oplus$ is too short to be detected.
The event in the bottom panel of Figure \ref{fig-lc-non} has a longer planetary signature due to a 10000 $M_\oplus$ planet. However, the planetary signature is missed because the there are no data points during the period of perturbation. 
\section{Statistics of observable microlensing events}
\label{sec-sta} 

By repeating the steps described in the previous section as illustrated in Figure \ref{fig_sche}, we conduct a Monte Carlo simulation of microlensing events and probe their detectability for each specified Galactic longitude and latitude, so that we obtain the expected number of microlensing events and planets.

{In the first four subsections, we calculate the number of detections of microlensing events. The yields of microlensing events for each Galactic coordinate per square degree during the survey duration $T_{\rm obs}$, $N_{\rm ML}(l,b)$, are derived by multiplying the number of source stars, $N_{\rm source}(l,b)$, the event rate, $\Gamma_{\rm source}(l,b)$, and the detection efficiency of microlensing events, $\epsilon_{\rm ML}(l,b)$,
\begin{equation}\label{eq-nml}
N_{\rm ML} (l,b) = \Gamma_{\rm source}N_{\rm source}T_{\rm obs}\epsilon_{\rm ML}.
\end{equation}
We show the distribution of $N_{\rm source}$ and $\Gamma_{\rm source}$ for each field at first in Figure \ref{fig_Starmap} and Figure \ref{fig_EventRate}. Then we show the results of the estimation of the detection efficiency and the number of detections of microlensing events as a function of field coordinate and observation cadence in Figure \ref{fig_DES} and Figure \ref{fig_ML_yields}.}

{In the last two subsections, we also calculate the number of detections of planets per square degree per year, $N_{\rm PL}(l,b)$ in Figure \ref{fig_PL_yields}, as follows,
\begin{align}\label{eq-npl}
&N_{\rm PL} (l,b)   \notag \\ 
&=\int_{a=0.3{\rm au}}^{a=30{\rm au}}\int_{M_p=0.1{M_\oplus}}^{M_p=10^5{M_\oplus}} N_{\rm ML} \epsilon_{\rm PL}f_p d\log(a)d\log(M_p).
\end{align}
where $\epsilon_{\rm PL}(l,b,a,M_p)$ is the detection efficiency of planets and $f_p[\log(a),\log (M_p)]$ is the cool-planet mass function.} 

We conduct our simulation for 875 fields over $-4.25^{\circ} < l < 4.5^{\circ}$ and $-3.25^{\circ} < b<3^{\circ}$ with a resolution of $0.25^{\circ}\times0.25^{\circ}$. The \citet{2020A&A...644A.140S} extinction map used in \texttt{genstars} has up to $0.0025^\circ \times 0.0025^\circ$ resolution. 
{To reduce the computational time without losing the extinction variation, the number of the sources and lenses in catalogs are reduced by a scaling factor $f_{\rm sim}$ in \texttt{genstars}. The source and lens catalogs for each grid are created by giving the grid size of $0.25^{\circ}\times 0.25^{\circ}$, where the extinction variation with the resolution of $0.0025^\circ \times 0.0025^\circ$ are taken into account. Then we use the scaling factor $f_{\rm sim}$ = 0.0032 that reduces uniformly to 0.0032 times the number of stars in the given grid. Along each grid and each observation cadence, we randomly generate one hundred thousand microlenisng events by using the source and lens catalogs.}


\subsection{Source star counts}\label{sec-star-counts}

\begin{figure}[htbp]
      \hspace*{-1cm}
      \includegraphics[scale=0.25]{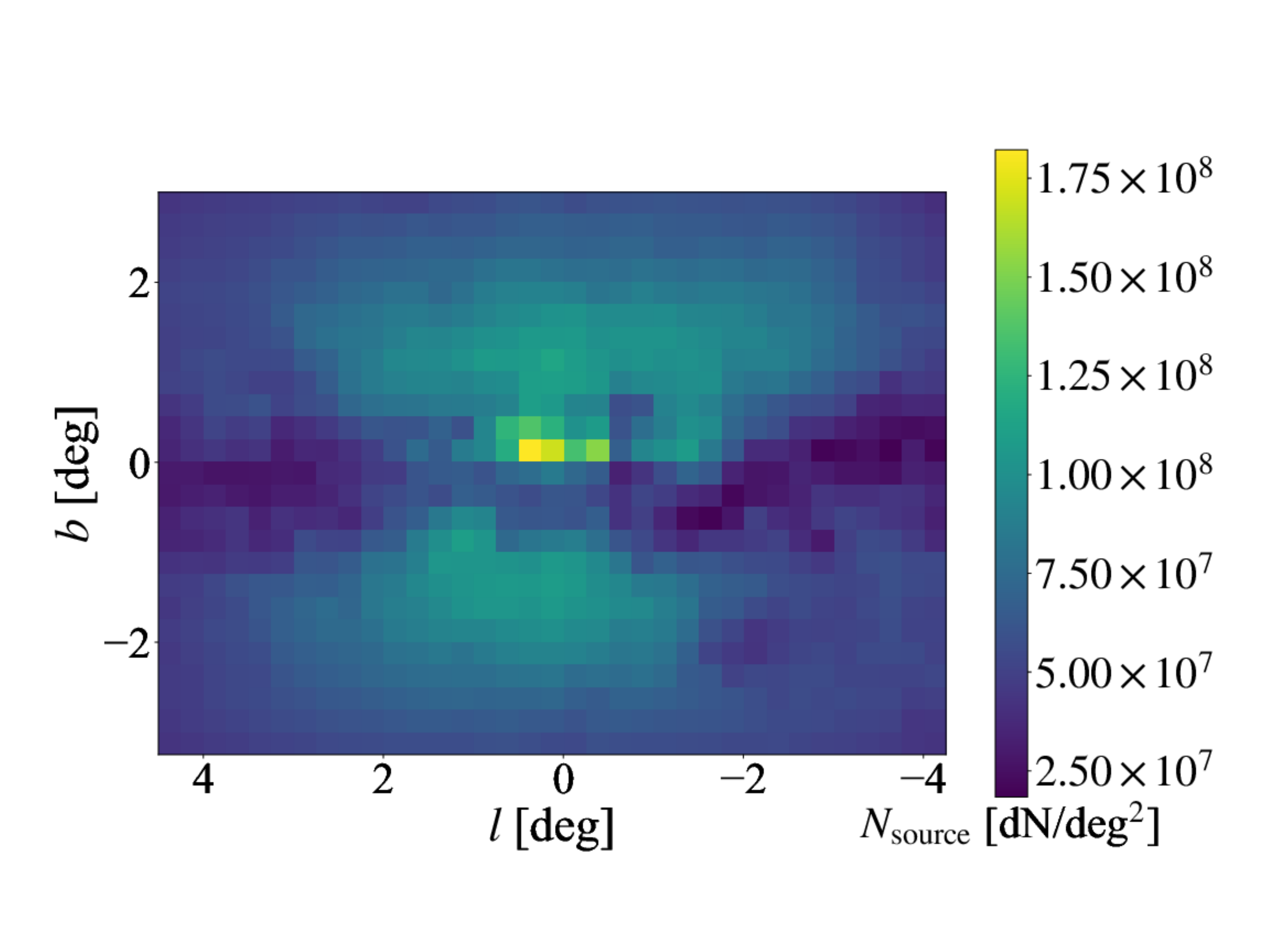}
  \caption{Map of star counts with $10.5<H_s<22$ mag, $N_{\rm source} (l, b)$, in our source catalogs generated by \texttt{genstars}. Most of stars in the region $|b|<0.5^{\circ}$ and $|l|<1.5^{\circ}$ belong to the NSD component, yielding high stellar density. However owing to the high extinction, the number of source is few in the Galactic center and the Galactic plane.}
  \label{fig_Starmap}
\end{figure}

Figure \ref{fig_Starmap} shows the KGM stellar density map for stars with $10.5 < H_S < 22$; $N_{\rm source} (l, b)$, calculated from the source catalogs.
The star counts per square degree, $N_{\rm source} (l,b)$, along the line of sight is calculated as,

\begin{equation}
    N_{\rm source}(l,b) = \frac{N_{\rm sim}}{f_{\rm sim}\delta\Omega_S},
\end{equation}
where $N_{\rm sim}$ is the number of source stars generated by \texttt{genstars}, $\delta\Omega_S =0.25^{\circ}\times0.25^{\circ}$ is the solid angle within which each source is drawn from \texttt{genstars}, and  $f_{\rm sim}= 0.0032$ is the scaling factor that we specified to limit the number of output stars by \texttt{genstars}.

Star counts depend on the combination of stellar number density and extinction.
Most of stars in the region $|b|<0.5^{\circ}$ and $|l|<1.5^{\circ}$ belong to the NSD component, yielding a relatively high stellar density. However, owing to high extinction, the number of sources is few in the Galactic center and the Galactic plane. {Therefore, according to Figure \ref{fig_Starmap}, the mean number of stars in the region $-0.75<b<0.5$ is $\sim 5.3\times 10^7$ stars per square degree, which is $\sim23\%$ and $\sim12\%$ lower than that in the region $-2.0<b<-0.75$ and $-3.25<b<-2.0$, respectively.}

\begin{figure}[ht]
      \includegraphics[scale=0.4]{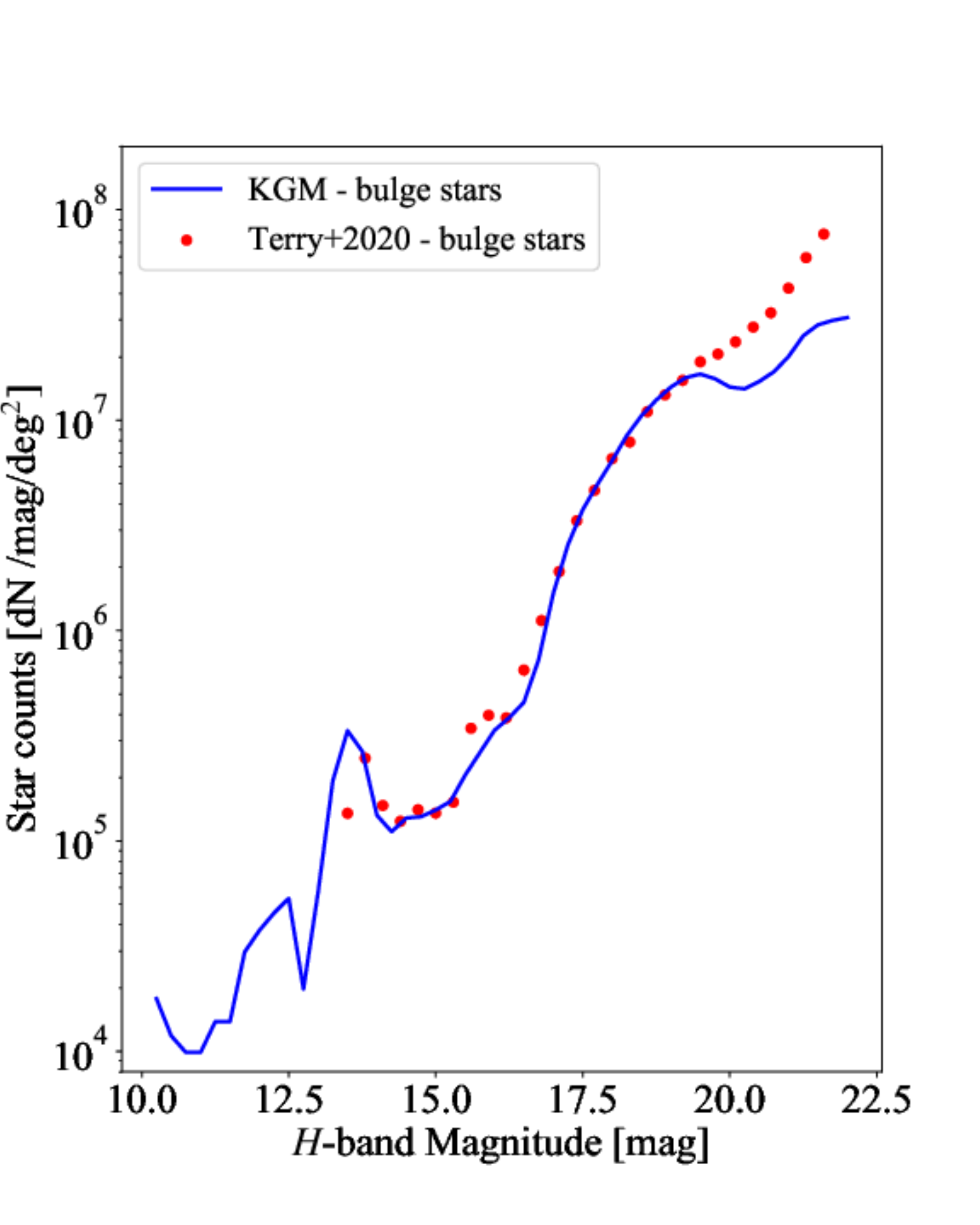}
  \caption{Comparison of star counts in Stanek window ($l,b=[0.25^{\circ},-2.15^{\circ}]$) in KGM (blue line) for the bulge population as a function of $H$-band magnitude to those by $HST$ observation in \citet{2020ApJ...889..126T} (red points). Stars with $H>19.5$ mag are underestimated in the Galactic model.}
  \label{fig_Starcounts}
\end{figure}

We also compare the bulge star counts by KGM with that by observation for validation. Figure \ref{fig_Starcounts} shows a comparison between luminosity functions in the Stanek window ($l, b=[0.25^{\circ}, -2.15^{\circ}]$) predicted by the KGM and as observed by the {\it Hubble Space Telescope} ($HST$) \citep{2020ApJ...889..126T}. \citet{2020ApJ...889..126T} distinguished between foreground stars and bulge stars by accurate measurement of the longitudinal proper motion. Although we should use same cut of the proper motion as \citet{2020ApJ...889..126T}, here we plot the counts for stars labeled bulge stars in the output catalog by \texttt{genstars}.

Figure \ref{fig_Starcounts} shows that stars with $H \simgt 19.5$ mag are underestimated in KGM. 
However, this discrepancy is not expected to affect simulation results for two reasons. First, at the Galactic center and in the Galactic plane ($|l|<2^{\circ}, |b|<1^{\circ}$), owing to the high extinction $A_H\footnote{We estimate $A_H$ using the \citet{2020A&A...644A.140S}'s $E(J - K_s)$ map and the \citet{2009ApJ...696.1407N}'s extinction law.}\sim 1.5-3.5$ compared to the extinction in the Stanek window $A_{H,{\rm stanek}}\sim0.68$, the underestimated faint stars are expected to be almost undetectable by PRIME even if the magnification is high. Second, at fields away from the Galactic center (e.g. $|l|<2^{\circ}, -2^{\circ}<b<-1^{\circ}$), although the extinction ($A_H\sim 0.4-1.0$) is almost the same as that at the Stanek window, we expect little effect on the total result because of the small percentage of detectable events with $H_S \simgt 19.5$ owing to the low detection efficiency for faint source stars.

\subsection{Event Rate}

\begin{figure}[t]
      \hspace*{-1cm}
      \includegraphics[scale=0.25]{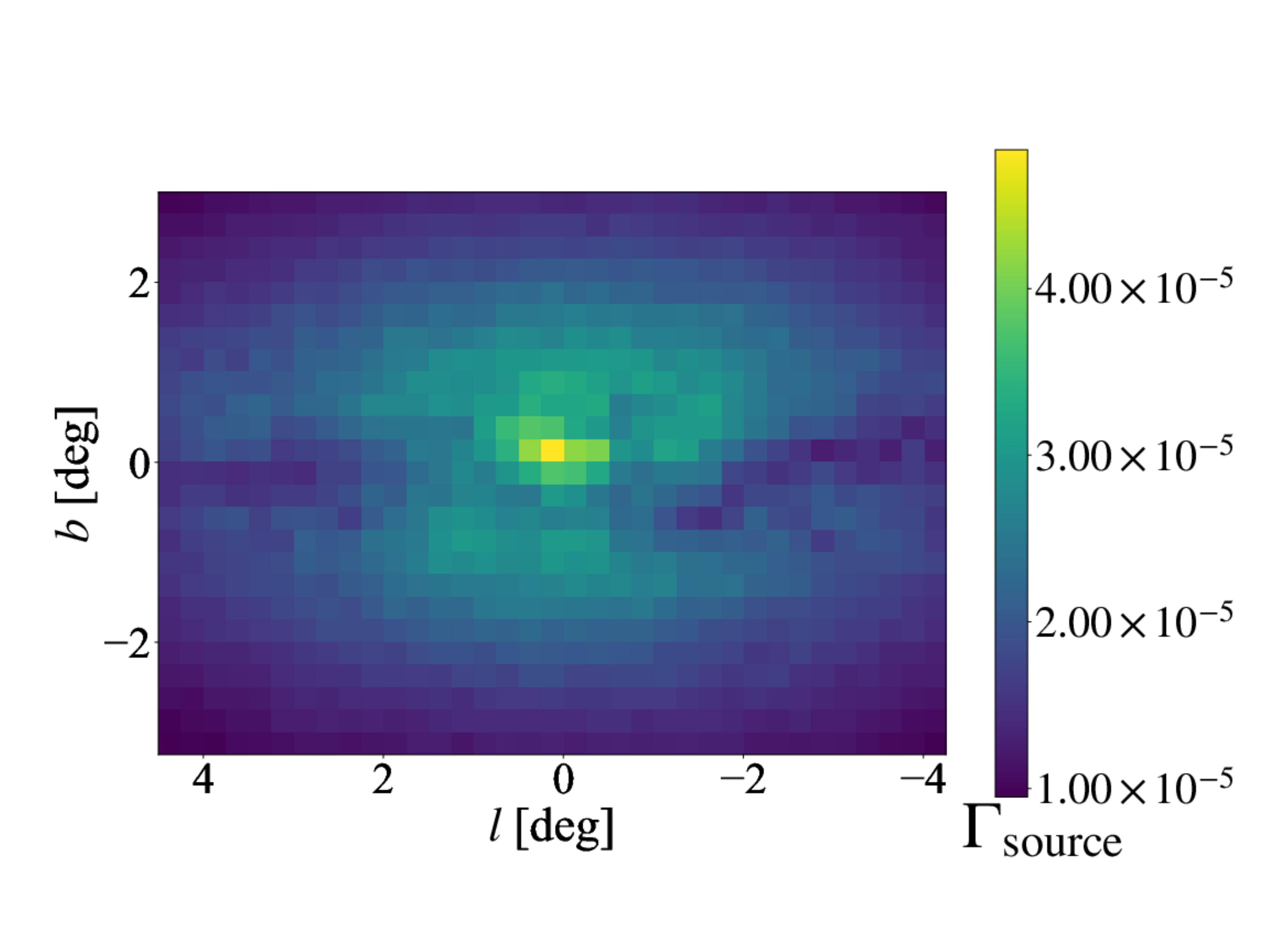}
  \caption{Map of event rate per source, $\Gamma_{\rm source}$, calculated using source and lens catalogs, generated by \texttt{genstars}. Event rate per source are mainly determined by the stellar density, so at the NSD region ($|b|<0.5^{\circ}, |l|<1.5^{\circ}$) the event rate is highest among other fields.}
  \label{fig_EventRate}
\end{figure}

\begin{figure}[ht]
      \hspace*{-1cm}
      \includegraphics[scale=0.4]{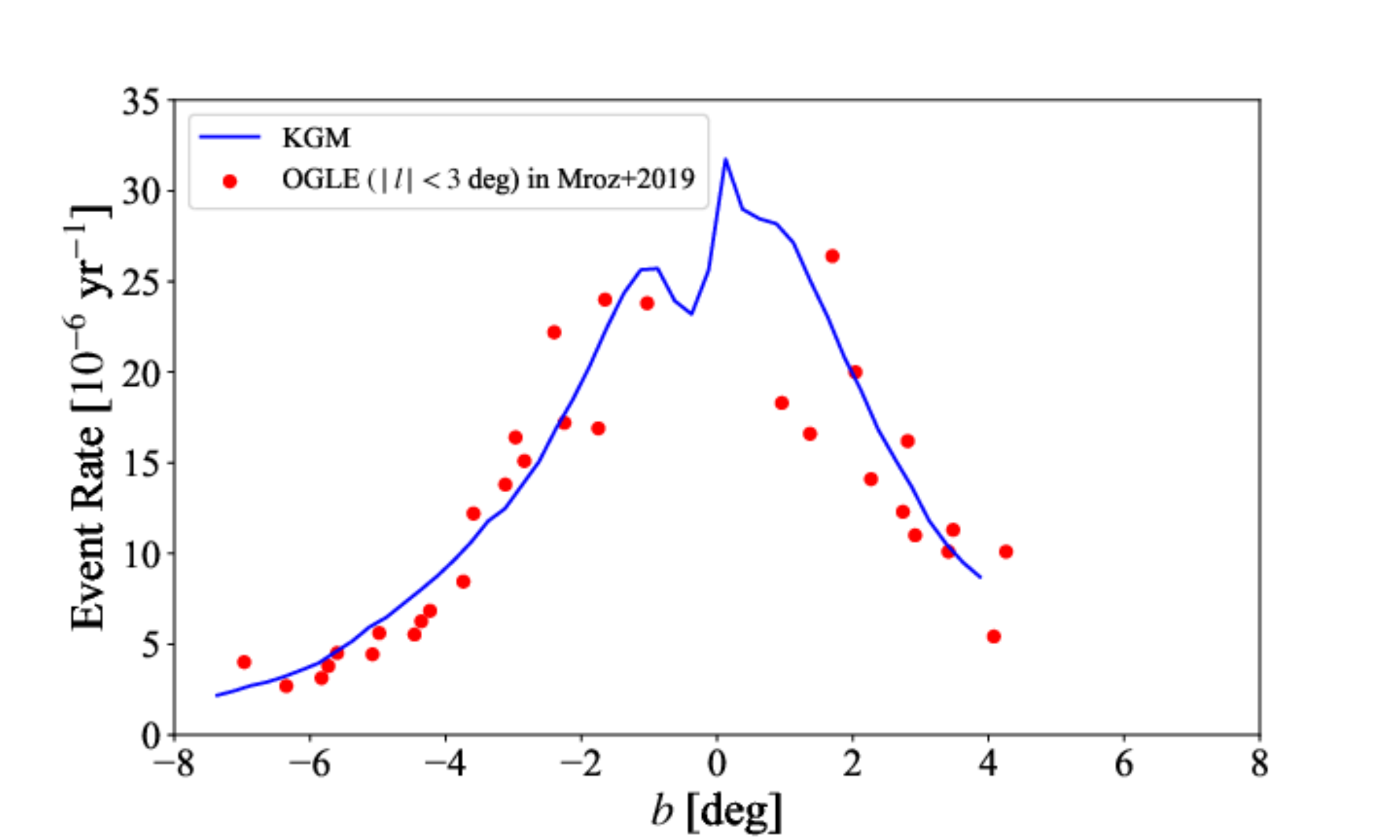}
  \caption{Comparison of the event rate per source calculated from the source and lens catalogs from \texttt{genstars} (blue line) with that measured in \citet{2019ApJS..244...29M} (red points). Due to the high extinction around the Galactic plane, there is no measurement of event rate at $|b|<1^{\circ}$ by OGLE. Outside of the Galactic plane, the two values of event rate are almost coincident.}
  \label{fig_EventRateComp}
\end{figure}

The microlensing event rate, $\Gamma_{\rm source} (l,b)$, is the probability that a source star is magnified by a foreground lens star per unit time.
The event rate per source is calculated via Monte Carlo integration of the event rate using source and lens catalogs as follows \citep{2016MNRAS.456.1666A,2013MNRAS.434....2P},
\begin{align}
& \Gamma_{\rm source} (l,b)   \notag \\ 
&=\frac{\Omega_{\rm los}}{f_{\rm sim}\delta\Omega_S}\frac{1}{N_{\rm sim}} \sum^{\rm sources} \left( \frac{1}{f_{\rm sim}\delta\Omega_l}\sum^{\rm Lenses}_{D_L<D_S} 2\theta_{\rm E} \mu_{\rm rel}\right),
\end{align}
where $\Omega_{\rm los}$ is the solid angle of each grid, and $\delta\Omega_S$ and $\delta\Omega_L$ are the solid angle of the source and lens catalogs, respectively. In our simulation, we use $\Omega_{\rm los} = \delta\Omega_S = \delta\Omega_L = 0.25^{\circ}\times0.25^{\circ}$.

Figure \ref{fig_EventRate} shows the KGM map of event rate per source, $\Gamma_{\rm source} (l,b)$, derived using our source and lens catalogs. 
According to Figure \ref{fig_EventRate}, at the NSD region ($|b|<0.5^{\circ}, |l|<1.5^{\circ}$) the event rate is highest among all fields. This is because $\Gamma_{\rm source} (l,b)$ is mainly determined by stellar density.
{The mean event rate per source in the region $-0.75<b<0.5$ is $\sim 2.5\times 10^{-5}$, which is $\sim 15\%$ and $\sim73\%$ higher than that in the region $-2.0<b<-0.75$ and $-3.25<b<-2.0$, respectively.}

Figure \ref{fig_EventRateComp} compares the model event rate values with the observational values by \citet{2019ApJS..244...29M}. \citet{2019ApJS..244...29M} shows the optical depth and event rate maps by using the largest sample of 8000 events from the optical survey of OGLE-IV during $2010-2017$. Owing to the high extinction around the Galactic center, there is no measurement of event rate at $|b|<1^{\circ}$ by OGLE. Outside of the Galactic plane, the two values of event rate are almost coincident, thus we conclude that there is no need of correction for the model event rate values as was done in \citet{2019ApJS..241....3P}.

\subsection{Detection efficiency for microlensing events}

We estimate the detection efficiencies for microlensing events, $\epsilon_{\rm ML} (l,b)$ along each line of sight of the inner Galactic bulge. 
Using the detection criteria described in Section \ref{sec-det-cri-ml}, detection efficiency of microlensing events, $\epsilon_{\rm ML} (l,b)$ is defined as the ratio of the number of detected events to the number of all simulated events and calculated as
\begin{equation}
\epsilon_{\rm ML} (l,b) = \frac{\Sigma_{i,{\rm microlensing}}\  2 \mu_{{\rm rel},i}\theta_{{\rm E},i}}{\Sigma_{i,{\rm all}}\  2\mu_{{\rm rel},i}\theta_{{\rm E},i}},
\end{equation}
where each event $i$ is weighted by its microlensing event rate ($\propto 2\mu_{{\rm rel},i}\theta_{{\rm E},i}$).  

\begin{figure*}[ht] 
 \begin{center}
  \includegraphics[scale=0.35]{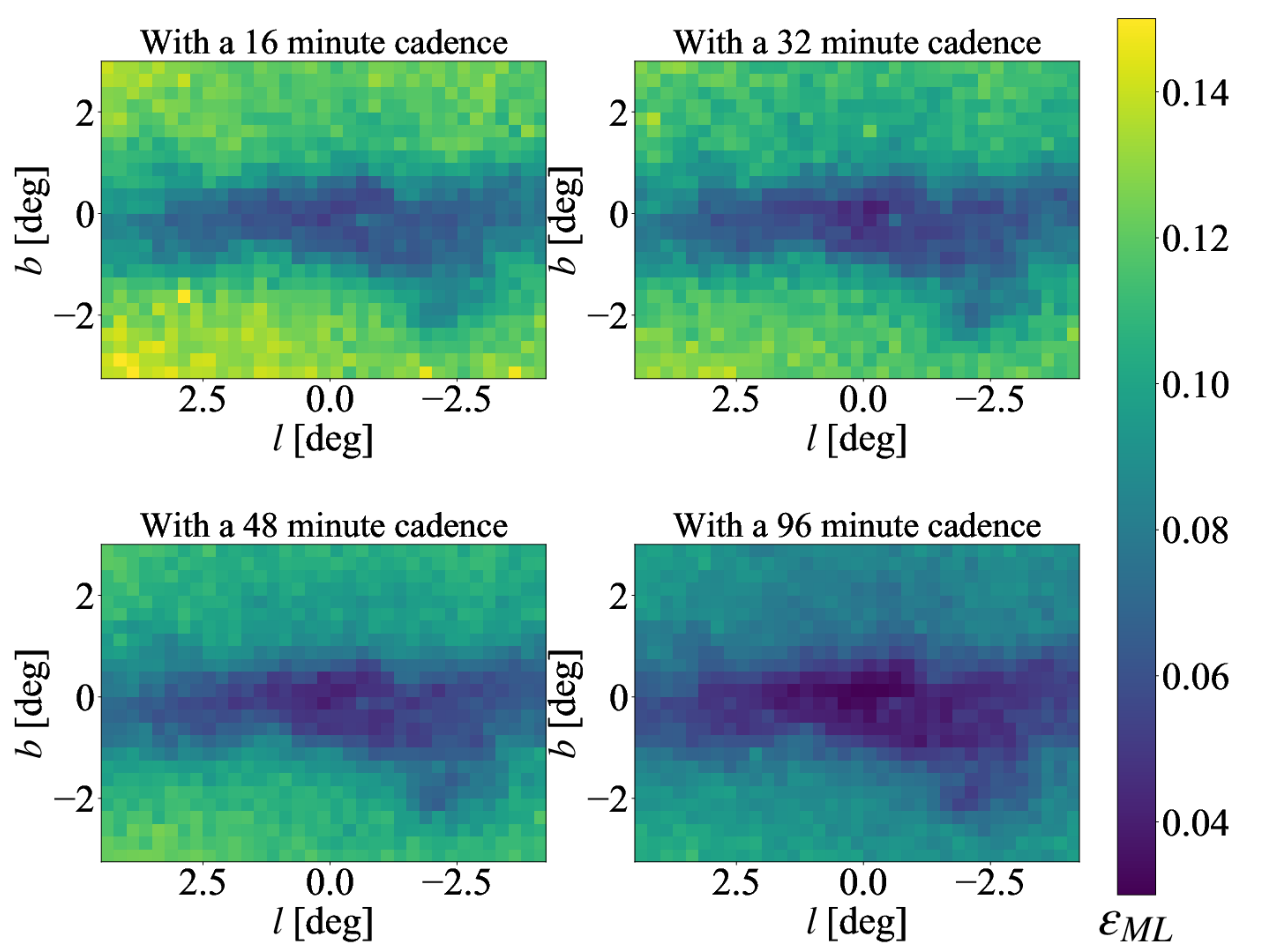}
  \caption{Mean detection efficiency of microlenisng events along each line of sight, $\epsilon_{\rm ML}(l,b)$. Each plot shows the detection efficiency for different cadences. With the same observation cadence, the detection efficiency is lower at the Galactic center than away from the Galactic center. See the text for an explanation of these trends. At the same field, the lower the observation cadence, the lower the detection efficiency.}
  \label{fig_DES}
 \end{center}
\end{figure*}

\begin{figure*}[ht]
 \begin{center}
 \hspace*{-1cm}
  \includegraphics[scale=0.2]{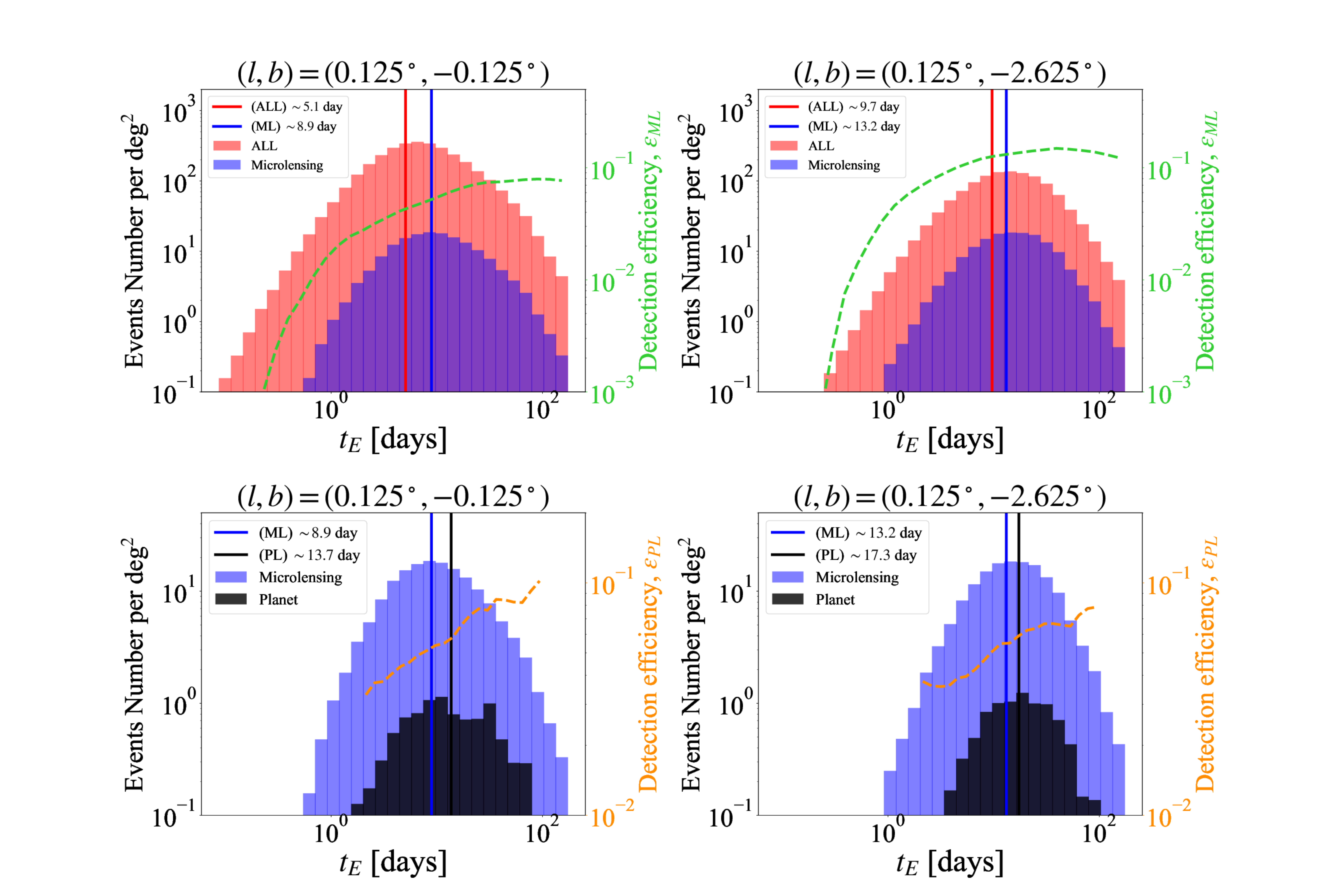}
  \caption{The Einstein ring crossing time, $t_{\rm E}$, distribution at $(l, b) = (0.125^{\circ}, -0.125^{\circ})$ (left panels) and at $(l, b) = (0.125^{\circ}, -2.625^{\circ})$ (right panels). Top panels show the distribution of all simulated events (red) and detected microlensing events (blue) with a 16 minute cadence by the assumed PRIME survey. Bottom panels show the distribution of detected microlensing events (blue) and detected planetary events (black). The vertical lines show the median value of each histogram. The dashed green and orange lines show the detection efficiency of microlensing events and planetary events depending on $t_{\rm E}$, respectively.} 
  \label{fig_tE_comp}
 \end{center}
\end{figure*}

\begin{figure*}[ht]
 \begin{center}
 \hspace*{-1cm}
  \includegraphics[scale=0.2]{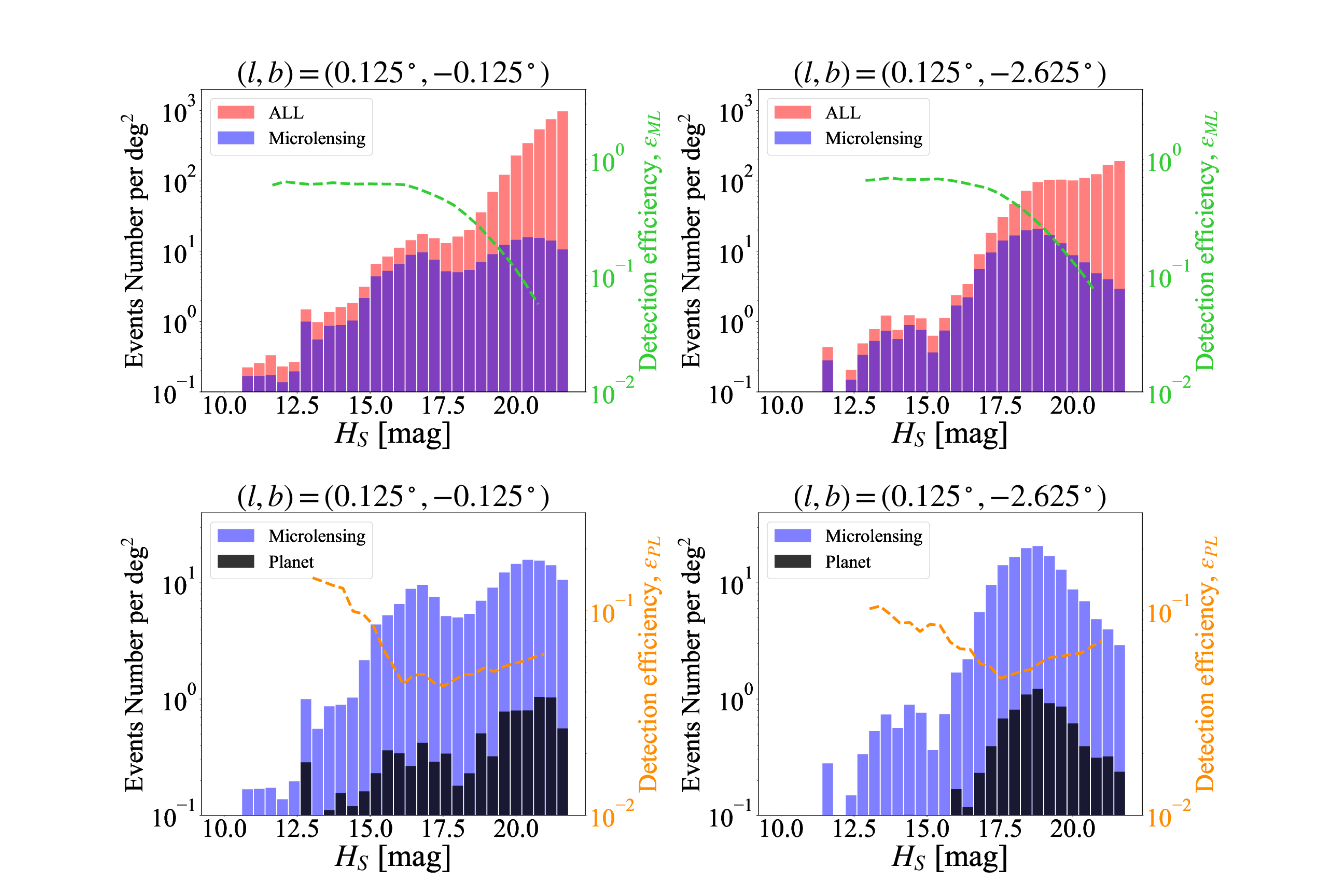}
  \caption{The source magnitude, $H_S$, distribution at $(l, b) = (0.125^{\circ}, -0.125^{\circ})$ (left panels) and at $(l, b) = (0.125^{\circ}, -2.625^{\circ})$ (right panels).  Top panels show the distribution of all simulated events (red) and detected microlensing events (blue) with a 16 minute cadence. Bottom panels show the distribution of detected microlensing events (blue) and detected planetary events (black). The dashed green and orange lines show the detection efficiency of microlensing events and planetary events depending on $H_S$, respectively.} 
  \label{fig_Hmag_comp}
 \end{center}
\end{figure*}

Figure \ref{fig_DES} shows the mean detection efficiencies of microlensing events along each line of sight with 16, 32, 48, and 96 minute cadences. At the same observation cadence, detection efficiency is lower at the Galactic center than away from the Galactic center. 
{The mean number of detection efficiencies with a 16 minute cadence in the region $-0.75<b<0.5$ is $\sim 0.07$, which is $\sim29\%$ and $\sim43\%$ lower than that in the region $-2.0<b<-0.75$ and $-3.25<b<-2.0$, respectively.}
There are two reasons why the mean detection efficiency of microlensing events is lower at the Galactic center. 
The first reason is the large fraction of short $t_{\rm E}$ events at the Galactic center. The top panels in Figure \ref{fig_tE_comp} show $t_{\rm E}$ distributions for all simulated events (red histogram) and detected events (blue histogram) at two Galactic coordinates. 
{The median value of $t_{\rm E}$ at $(l, b) = (0.125^{\circ}, -0.125^{\circ})$, is $\sim 5.1$ days, which is smaller than $\sim 9.7$ days at $(l, b) = (0.125^{\circ}, -2.625^{\circ})$}, because the majority of events toward the former direction comprise a source and a lens located in the bulge, yielding the small lens-source relative parallax, $\pi_{\rm rel}$ and small angular Einstein ring radius $\theta_{\rm E}$ (Equations (\ref{eq-thetae}) and (\ref{eq-te})). Microlensing events with short $t_{\rm E}$ are detected less efficiently by the survey as indicated by the green lines in Figure \ref{fig_tE_comp}. Therefore the mean detection efficiency, $\epsilon_{\rm ML}$, at $(l, b) = (0.125^{\circ}, -0.125^{\circ})$, is lower than that at $(l, b) = (0.125^{\circ}, -2.625^{\circ})$.
The second reason is the large fraction of faint stars owing to the high extinction at the Galactic center.
The top panels in Figure \ref{fig_Hmag_comp} show the luminosity functions for both the all simulated events (red histogram) and detected events (blue histogram) in the same Galactic coordinates as Figure \ref{fig_tE_comp}. 
The estimated extinction values are $A_H\sim 4.4$ and $A_H\sim 0.7$ for at $(l, b) = (0.125^{\circ}, -0.125^{\circ})$ and at $(l, b) = (0.125^{\circ}, -2.625^{\circ})$, respectively. The detection efficiency as a function of $H_S$ is lower for faint stars than for bright stars as indicated by the green lines in Figure \ref{fig_Hmag_comp}. The fraction of faint sources with $H_S>17.5$ in all events, which are lower $\epsilon_{\rm ML}$, is $\sim30\%$ and $\sim6\%$, at $(l, b) = (0.125^{\circ}, -0.125^{\circ})$ and $(l, b) = (0.125^{\circ}, -2.625^{\circ})$, respectively. Therefore, owing to high extinction, the large fraction of faint stars, whose detection efficiency is low, also results in low mean detection efficiency at the Galactic center.

Figure \ref{fig_DES} also shows that, at the same field, the lower the cadence, the lower the detection efficiency. Compared to the mean detection efficiency in the same region with a 16 minute cadence, the detection efficiencies are $\sim 9\%,\ 17\%,\ 33\%$ lower with 32, 48, 96 minute cadences, respectively.

\begin{figure}[ht]
 \begin{center}
 \hspace*{-1cm}
  \includegraphics[scale=0.20]{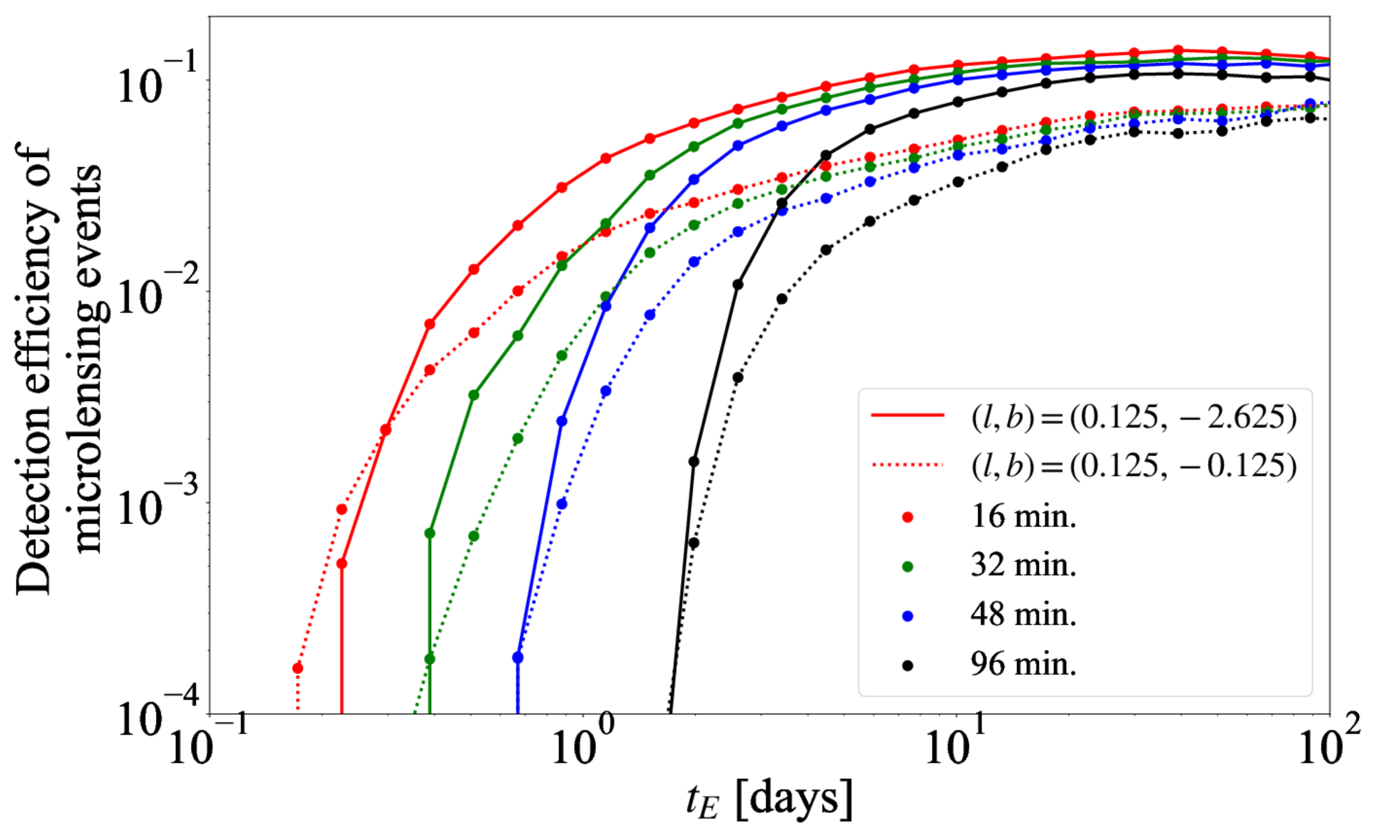}
  \caption{The detection efficiency of microlensing events depending on $t_{\rm E}$. The solid and dotted lines show the detection efficiency away from the Galactic center, $(l,b)=(0.125,-2.625)$, and at the Galactic center, $(l,b)=(0.125, -0.125)$, respectively. The detection efficiency with 16, 32, 48, and 96 minute cadences are shown in red, green, blue, and black, respectively.} 
  \label{fig_tE-DES}
 \end{center}
\end{figure}

In Figure \ref{fig_tE-DES}, we plot the detection efficiency of microlensing events depending on the Einstein crossing time, ${t_{\rm E}}$. As expected, the detection efficiency becomes lower near the Galactic center and/or with lower cadence.  It is difficult to detect microlensing events with $t_{\rm E} \simlt $ 0.3, 0.6, 1, and 3 days when the observation cadence is 16, 32, 48, and 96 minutes, respectively.

\clearpage

\subsection{The Number of Detected Microlensing events} \label{sec-yield-pr}
\begin{figure*}[ht]
 \begin{center}
  \includegraphics[scale=0.35]{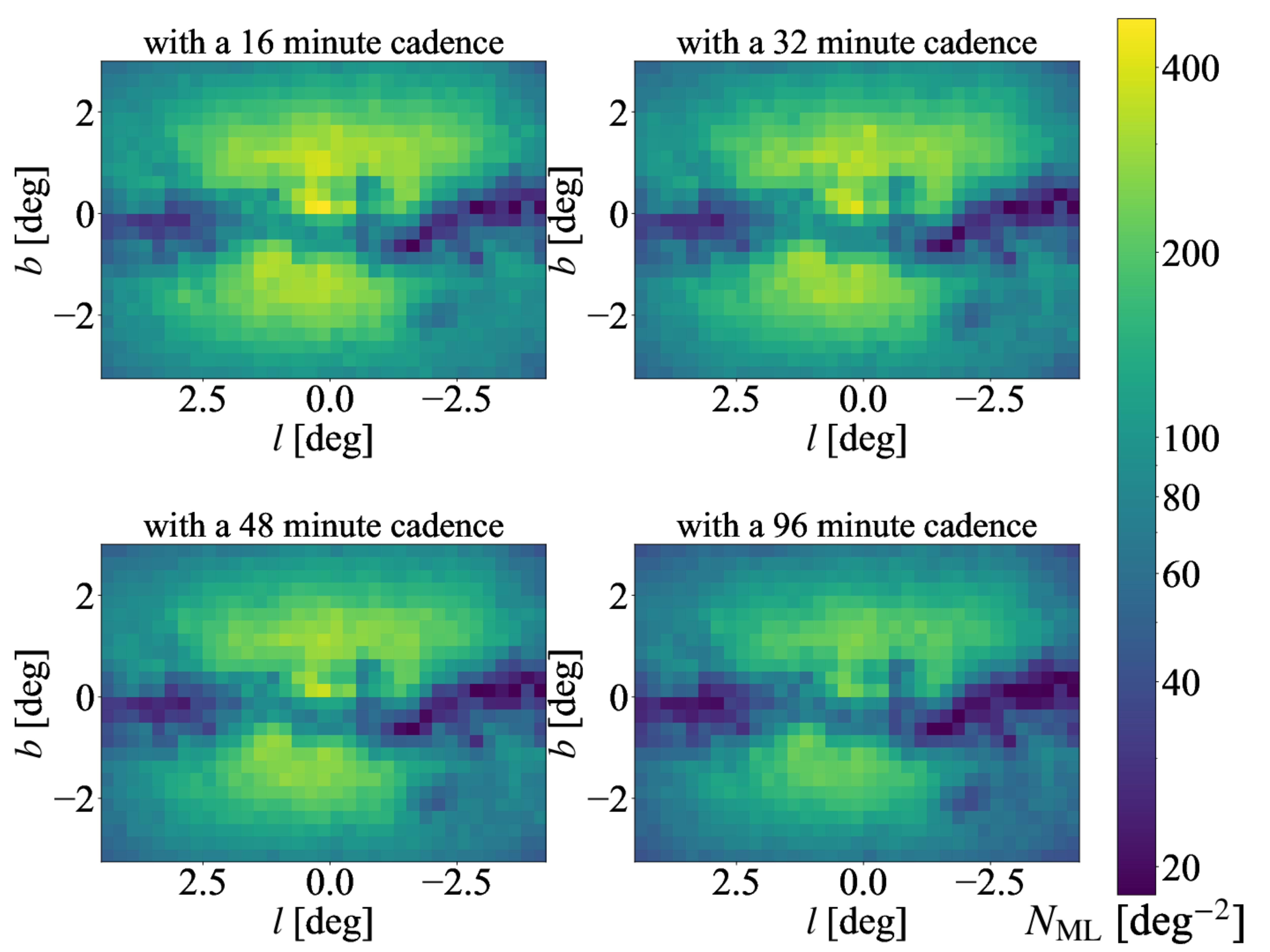}
  \caption{Microlensing detection maps along each line of sight. Each plot shows the number of detections with 16, 32, 48, and 96 minute cadences. This figure is obtained by multiplying star counts, $N_{\rm source}$, (Figure \ref{fig_Starmap}), event rate, $\Gamma_{\rm source}$, (Figure \ref{fig_EventRate}) and mean detection efficiency of microlenising events, $\epsilon_{\rm ML}$, (Figure \ref{fig_DES}).}
  \label{fig_ML_yields}
 \end{center}
\end{figure*}

Figure \ref{fig_ML_yields} shows the yields of microlensing events for each Galactic coordinate per square degree for one year, $N_{\rm ML} (l,b)$, calculated by Equation (\ref{eq-nml}).
According to Figure \ref{fig_ML_yields}, 
{the mean number of microlensing yields with a 16 minute cadence in the region $-0.75<b<0.5$ is $\sim 93$ events per square degree, which is $\sim41\%$ and $\sim18\%$ lower than that in the region $-2.0<b<-0.75$ and $-3.25<b<-2.0$, respectively.}
Compared to the microlensing yields in the same region with a 16 minute cadence, the yields are $\sim 10\%,\ 18\%,\ 35\%$ lower with 32, 48, 96 minute cadences, respectively.

\subsection{Detection efficiency for planetary signatures}
We also estimate the detection efficiencies of the planetary signatures $\epsilon_{\rm PL} (l,b,a,M_p)$ along each line of sight. Following the detection criteria of planetary signatures described in Section \ref{sec-det-cri-pl}, detection efficiency of a planetary signature is defined as the ratio of the number of detected planetary events to the number of detected events as microlensing
\begin{equation}
\epsilon_{\rm PL} (l,b,a,M_p) = \frac{\Sigma_{i,{\rm planet}}\  2 \mu_{{\rm rel},i}\theta_{{\rm E},i}}{\Sigma_{i,{\rm microlensing}} \ 2 \mu_{{\rm rel},i}\theta_{{\rm E},i} }.
\end{equation}

\begin{figure}[h]
 \begin{center}
  \includegraphics[scale=0.22]{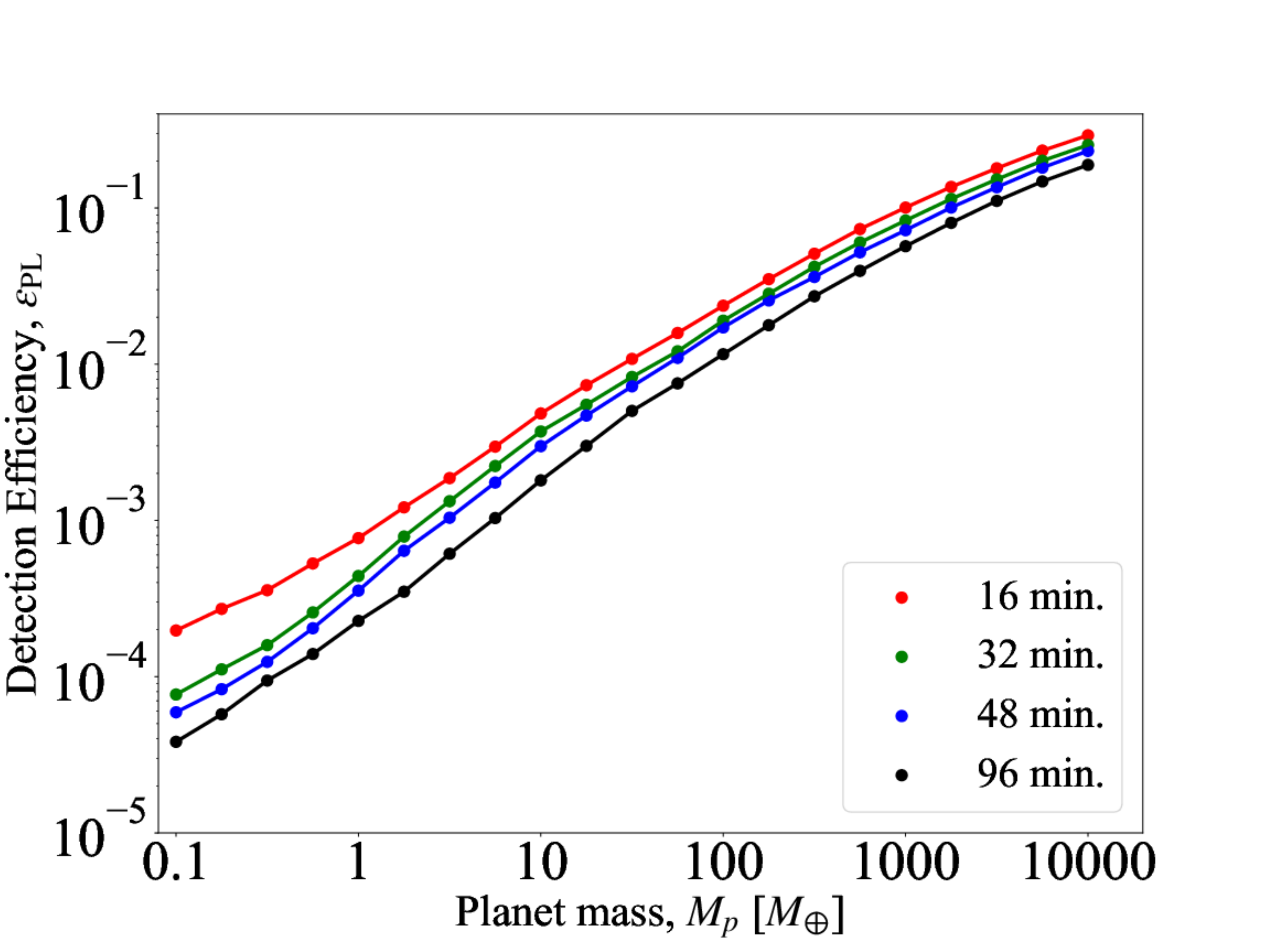}
  \caption{The detection efficiency of planetary signatures, $\epsilon_{\rm PL}(M_p)$ depending on planet mass, which are obtained by taking the average of all 875 fields and are summed across semi-major axis, $0.3<a<30$ au. Red, green, blue, and black color plots shows detection efficiency with 16, 32, 48, and 96 minute cadences, respectively.}
  \label{fig_DEP}
 \end{center}
\end{figure}

Figure \ref{fig_DEP} shows the detection efficiency of planetary signatures, $\epsilon_{\rm PL} (M_p)$, as a function of planet mass, which are obtained by averaging over all 875 fields and are summed across semi-major axis, $0.3<a<30$ au. With a 16 minute cadence, the detection efficiencies of Jupiter mass planet, Neptune mass, and Earth mass planet are $\sim 0.05$, $\sim 0.007$, and $\sim 0.0006$, respectively. Compared to the detection efficiency with a 16 minute cadence, the detection efficiency is $\sim 15-20\%$, $\sim 30-50\%$, and $\sim 50-70\%$ lower with 32, 48, and 96 minute cadences, respectively. In addition, the degree of decrease in detection efficiency with observation cadence is greater for low-mass planets.

We note that detection efficiency of the planetary signature can be regarded as almost the same over all fields simulated, owing to the combination of $t_{\rm E}$ distributions and luminosity functions. 
Firstly, at the Galactic center, the fraction of short $t_{\rm E}$ events is larger than that away from the Galactic center.
The bottom panels in Figure \ref{fig_tE_comp} show $t_{\rm E}$ distributions for both the detected microlensing events (blue histogram) and detected planetary events (black histogram) at two Galactic coordinates. 
{The median values of $t_{\rm E}$ for microlensing events at $(l, b) = (0.125^{\circ}, -0.125^{\circ})$, is $\sim 8.9$ days, which is smaller than $\sim 13.2$ days at $(l, b) = (0.125^{\circ}, -2.625^{\circ})$.}
Planetary events with short $t_{\rm E}$ are detected less efficiently by the survey, see the lines in Figure \ref{fig_tE_comp} describing $\epsilon_{\rm PL}$, as well as the detection efficiency of microlensing events, $\epsilon_{\rm ML}$.
Secondly, the fraction of bright stars at $(l, b) = (0.125^{\circ}, -0.125^{\circ})$ is larger than that at $(l, b) = (0.125^{\circ}, -2.625^{\circ})$.
The bottom panels in Figure \ref{fig_Hmag_comp} show the luminosity functions for both the detected microlensing events (blue histogram) and detected planetary events (black histogram). 
The detection efficiency of planetary signatures, $\epsilon_{\rm PL}$ as a function of $H_S$ changes little for faint stars with $H_S > 16$, but are higher for bright stars with $H_S < 16$ as indicated by the lines in Figure \ref{fig_Hmag_comp}. The fraction of bright sources with $H_S<16$ in microlensing events, which are higher $\epsilon_{\rm PL}$, is $\sim20\%$ and $\sim7\%$, at $(l, b) = (0.125^{\circ}, -0.125^{\circ})$ and $(l, b) = (0.125^{\circ}, -2.625^{\circ})$, respectively.
Therefore, {the dependence of $\epsilon_{\rm PL}$ on Galactic coordinates is minimized} by the combination of the large fraction of short $t_{\rm E}$ events, which work to decrease mean detection efficiency, and the large fraction of bright stars, which work to increase mean detection efficiency, in microlenisng events at the Galactic center.

\subsection{The Number of Detected Planets}\label{sec-sta-dp}

\begin{figure*}[ht]
 \begin{center}
  \includegraphics[scale=0.35]{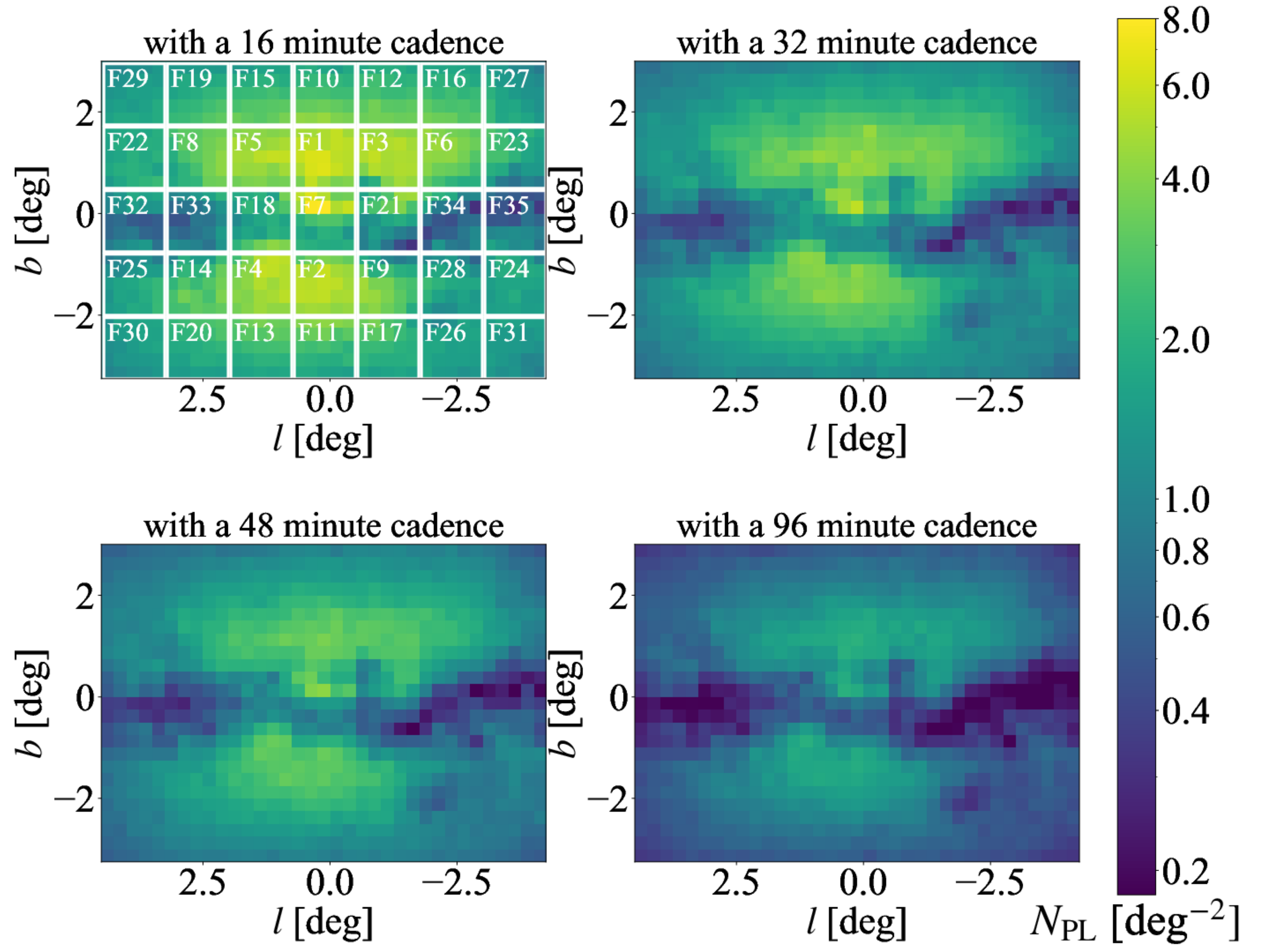}
  \caption{{Planet detection maps along each line of sight. Each plot shows the number of detections with 16, 32, 48, and 96 minute cadences. This figure is obtained by multiplying the number of microlensing detections, $N_{\rm ML}(l,b)$, (Figure \ref{fig_ML_yields}) and the mean detection efficiency of planets, $\epsilon_{\rm PL}$, which is obtained by the averaging over all fields, over mass of $0.1<M_p<10^5M_\oplus$, and semi-major axis of $0.3<a<30$ au and corrected by a modified cool-planet frequency based on \citet{2019ApJS..241....3P}. The planet detection map with a 16 minute cadence (upper left panel) is used to determine the order of the observation fields to in Section \ref{sec-str-ss}. Each white square shows a 1.45 deg$^2$ FOV field. The field numbers are ranked by the expected number of planet detections summed across each square.}}
  \label{fig_PL_yields}
 \end{center}
\end{figure*}

We calculate the number of the detectable planets per square degree per year, $N_{\rm PL}(l,b)$, by Equation (\ref{eq-npl}).
We use {the \citet{2012Natur.481..167C} mass function of planets beyond snow-line as modified by \citet{2019ApJS..241....3P},} which shows planet frequency per decade of mass and semi-major axis by using planets detected via microlensing. Because \citet{2012Natur.481..167C} {did not detect any planets} with a mass less than 5 $M_\oplus$, we decided to use a constant value, $\sim$ two planets per dex$^2$, below 5 $M_\oplus$ following \citet{2014ApJ...794...52H} and \citet{2013MNRAS.434....2P,2019ApJS..241....3P}. The mass function finally used can be stated as,
\begin{align}\label{eq-fp}
& f_p\lbrack \log(a),\log(M_p)\rbrack \equiv \frac{d^2N}{d \log(a)d \log(M_p)}  \notag \\ 
&=
  \begin{cases}
    \displaystyle {0.24 \ {\rm dex}^{-2} \left( \frac{M_p}{95M_{\oplus}}\right)^{-0.73}} & {\rm if}\ M_p \geq 5 M_\oplus, \\
    2 \ {\rm dex}^{-2}                 & {\rm if}\ M_p < 5 M_\oplus. \\
  \end{cases}
\end{align}

Figure \ref{fig_PL_yields} shows the planet detection maps computed using Equation (\ref{eq-npl}) along each line of sight.
According to Figure \ref{fig_PL_yields}, 
{the mean number of planets {detected} with a 16 minute cadence in the region $-0.75<b<0.5$ is $\sim 1.6$ events per square degree, which is $\sim41\%$ and $\sim18\%$ lower than that in the region $-2.0<b<-0.75$ and $-3.25<b<-2.0$, respectively.}
Compared to the planet {detections} in the same region with a 16 minute cadence, the yields are $\sim 31\%,\ 46\%,\ 70\%$ lower with 32, 48, 96 minute cadences, respectively.

The planet detection map with a 16 minute cadence (upper left panel in Figure \ref{fig_PL_yields}) is used to determine the order of the observation fields in the next section. The field numbers are ranked by the high expectation number of planet detections summed across each PRIME FOV.

{We investigate the impact of assuming other planet frequencies via microlensing as given in \citet{2016ApJ...833..145S} and \citet{2016MNRAS.457.4089S}. Figure 21 in \citet{2019ApJS..241....3P} shows a comparison of modified planet frequency based on \citet{2012Natur.481..167C} to the latest measurements of mass-ratio function by microlensing surveys \citep{2016ApJ...833..145S,2016MNRAS.457.4089S}. They assumed a $0.5M_\odot$ host star to convert mass-ratio to planet mass. The frequencies of low-mass planets ($M_p\simlt 30M_\oplus$) obtained in \citet{2016ApJ...833..145S} are lower than the modified planet frequency, which suggests lower yields of low-mass planets. 
However, the frequency of the Earth-mass planets is still not well understood owing to the lack of low-mass planets in the statistical analyses. 
The frequencies of high-mass planets ($3000<M_p/M_\oplus<10000$) obtained in \citet{2016MNRAS.457.4089S} are higher than the modified planet frequency, which suggests that the modified planet distributions underestimate planet yields for high-mass planets.}

\clearpage
\section{Observation Strategies and yields}\label{sec-str}

{Now that we have the expected number of microlensing events and planets as a function of Galactic coordinate and observation cadence, we are finally ready for discussing the PRIME survey strategy. }
{In this section, we define four observation strategies and calculate both microlensing yields and planet yields depending on each observation strategy.}

\subsection{Observation fields and strategies}\label{sec-str-ss}
{We divide our simulation fields ($|b| \simlt 2^\circ$, $|l| \simlt 4^\circ$) into 35 observation fields according to the size of the PRIME FOV and calculate the total number of planets expected to be detected in each observation field.
Then the observation field numbers are ranked in order of these total number of detections (upper left panel in Figure \ref{fig_PL_yields}).
Because the number of observation fields we can observe is determined by the observation cadence, we define four strategies as following and compare the planet yields among these four strategies:}

\begin{figure*}[ht]
  \begin{center}
      \includegraphics[scale=0.35]{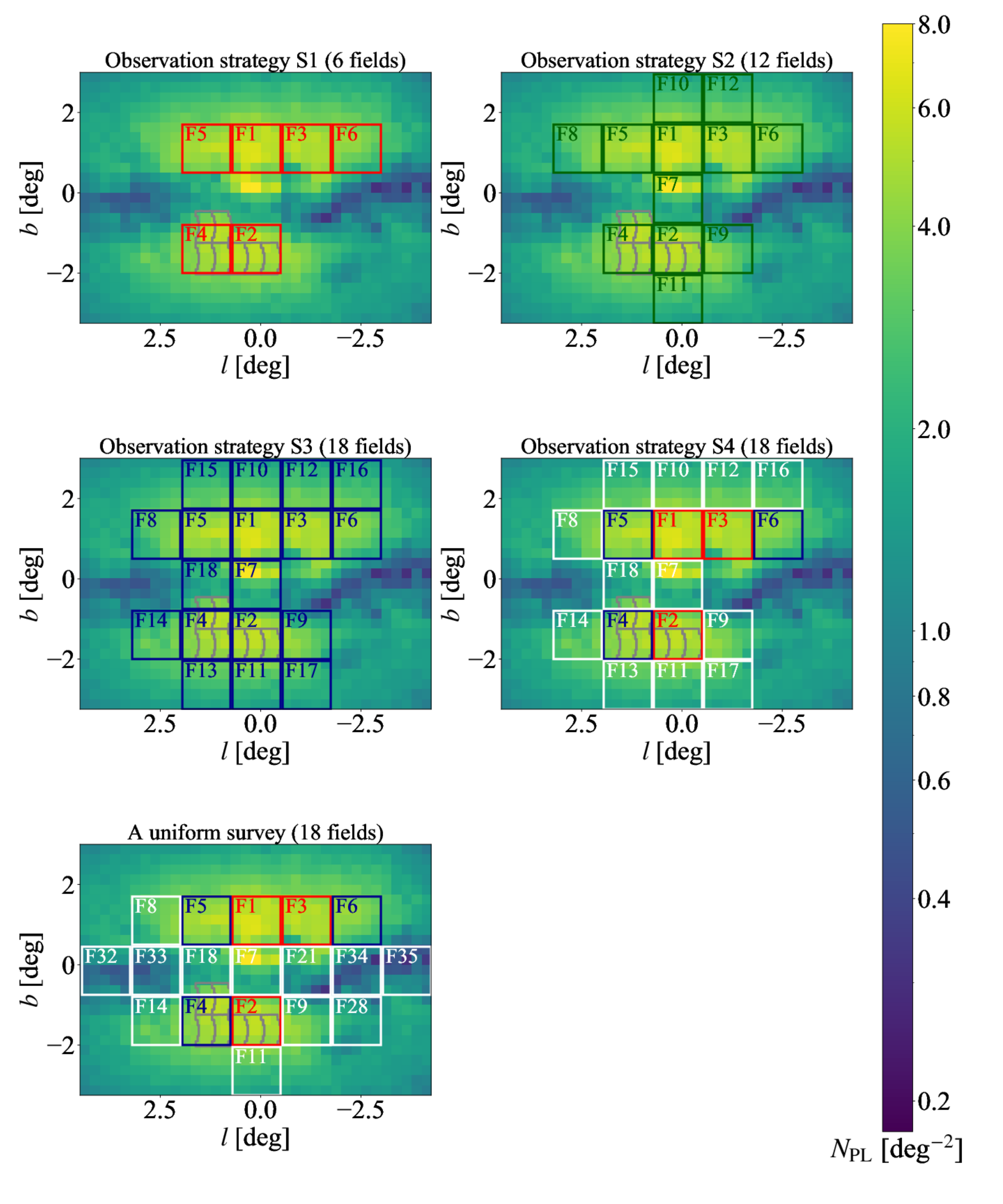}
  \end{center}
  \caption{Field locations for the PRIME microlensing survey for each observation strategy considered in this work, {plotted over the planet detection map with a 16 minute cadence. Top and middle panels show the observation strategies, S1--S4 described in Section \ref{sec-str-ss}.
  The bottom panel shows the spatially uniform survey including the Galactic center and the Galactic plane described in Section \ref{sec-dis-opt}.
  The field numbers are ranked by their expectation of planet detections (Figure \ref{fig_PL_yields})}. Each square shows a 1.45 ${\rm deg^{2}}$ FOV field, where the red, green, blue, and white indicate the cadences of 16, 32, 48, and 96 minutes, respectively. The gray region shows the assumed field placement for $Roman$ microlensing survey \citep{2019ApJS..241....3P}.}
  \label{fig_AHmap}
\end{figure*}

\begin{description}
    \item[S1] 6 fields (F1--F6) with a 16 minute cadence
    \item[S2] 12 fields (F1--F12) with a 32 minute cadence
    \item[S3] 18 fields (F1--F18) with a 48 minute cadence
    \item[S4] 18 fields (F1--F18) with a hybrid cadence (16min cadence for F1--F3, 48min cadence for F4--6, 96min cadence for the other 12 fields),
\end{description}
where we assumed that it takes 160 secs in total to observe a field (exposure + overheads) to calculate the cadence.
Figure \ref{fig_AHmap} shows all the 18 fields (F1--F18) considered here as well as which fields are observed by each strategy.
As shown in the figure, the S1, S2, and S3 strategies each have different survey regions and monitor all the fields in each region equally. The S4 strategy has the same survey region as S3, but each field is monitored with different cadence. We call S4 a hybrid strategy.

We consider these different strategies because there is a trade-off between the number of fields and frequency of observations. On the one hand, an increase of the number of fields allows us to monitor more sources, which will yield a lot of microlensing events. On the other hand, a higher cadence observation has a higher sensitivity to low-mass planets, because the timescales of the planetary signature scales with $\sqrt{q}t_{\rm E}$. 
The typical timescales of planetary signatures for Jupiter-mass planets and Earth-mass planets are a few days and a few hours, respectively.
Thus high cadence observations are required in order to detect Earth-mass planets, and it is unclear which strategy yields planet discoveries most efficiently including small mass planets without doing a simulation.
{However the following concerns caused by observations with a lower cadence are not considered in this paper. Lower cadence observations make it more difficult to measure the source radius crossing time, $\theta_*$($\equiv \rho t_{\rm E}$), and therefore $\theta_{\rm E}$. So it is more challenging to measure host and planet masses either by a combination of $\theta_{\rm E}$ and $\pi_{\rm E}$ measurements (as in \citealp{2011ApJ...741...22M}) or with the color dependent centroid shift \citep{2006ApJ...647L.171B,2009ApJ...695..970D}.}

Note that this paper is primarily concerned with the search for an optimal observation strategy with the goal of increasing planet yields to measure the planet frequency in the inner Galactic bulge. However we will discuss other observation strategies in Section \ref{sec-dis-opt}, including a uniform survey that monitors a large contiguous area around the inner Galactic bulge, in order to measure the NIR event rate map to help optimize the choice of $Roman$ microlenisng survey fields.

\begin{figure*}[ht]
 \begin{center}
 \hspace*{-0.5cm}
  \includegraphics[scale=0.20]{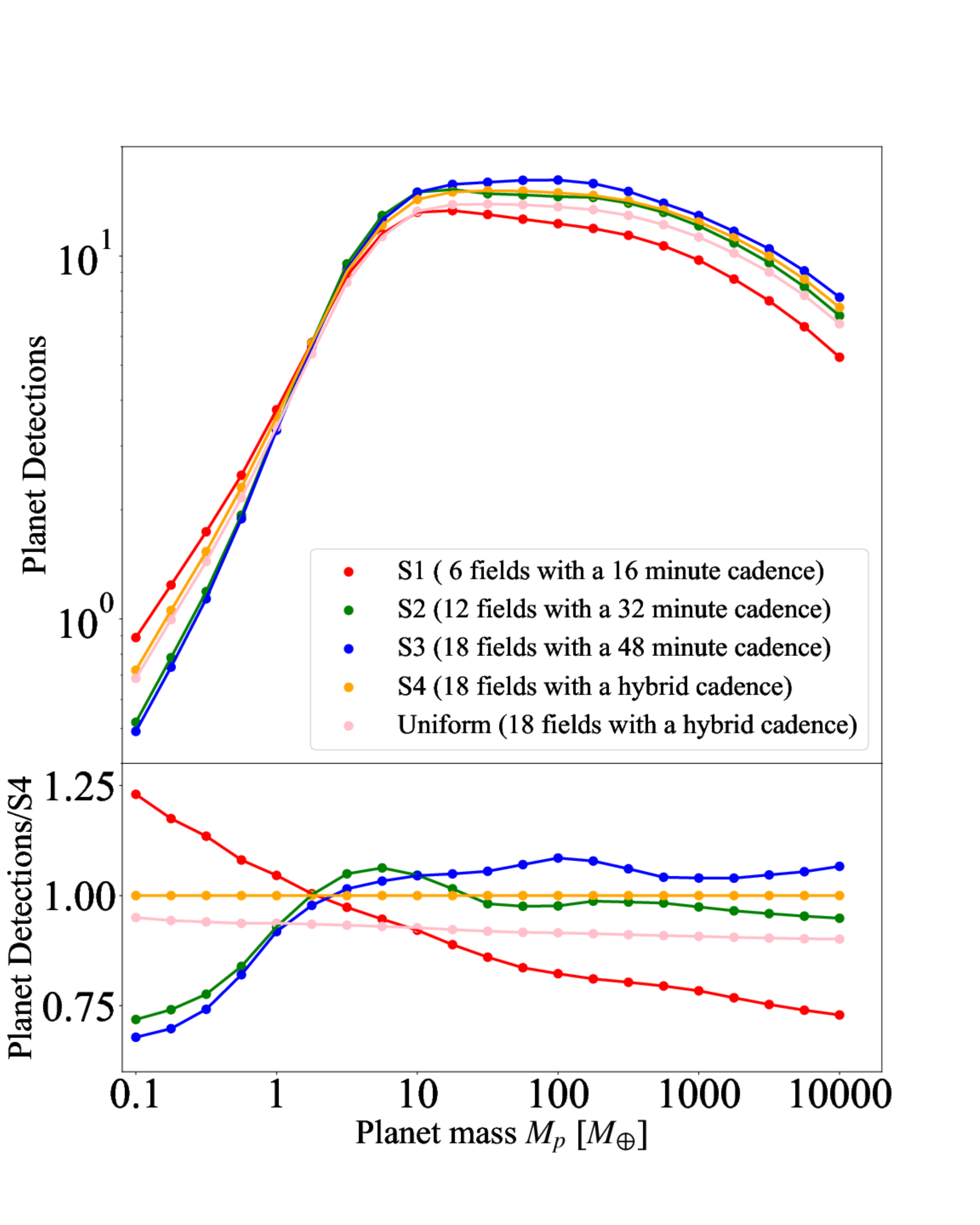}
  \caption{Upper panel shows the number of planet detections per dex as a function of planet mass, $M_p$. These plots are obtained by integrating over the semi-major axis $0.3<a<30$ au and over the survey area ($8.7 - 26.2$ deg$^2$) shown in Figure \ref{fig_AHmap}, assuming the \citet{2012Natur.481..167C} mass function as modified by \citet{2019ApJS..241....3P}. The red, green, blue, and orange plots show the detection rate for the observation strategy S1, S2, S3, and S4 described in Section \ref{sec-str-ss}. {The pink plot shows the detections when we conduct a spatially uniform survey described in Section \ref{sec-dis-opt}.}
  The lower panel shows the detections of each strategy relative to that of S4.}
  \label{fig_Planet_yields}
 \end{center}
\end{figure*}

\subsection{Yields}\label{sec-yield}
\begin{table*}
\centering
\caption{Best-estimate Planet Yields per year by the PRIME microlensing survey}
\label{Table_planet_log}
\begin{threeparttable}
 \begin{tabular}{cccccccc}
   \hline \hline
   Strategy & S1 & S2 & S3 & S4 & Uniform \\
   Total field number & 6 & 12 & 18 & 18 & 18\\
   Area(deg$^2$) & 8.7 & 17.5 & 26.2 & 26.2 &26.2\\
   Mass($M_\oplus$) & & & & & \\
   \hline
   $0.1<M_p\leq1.0$    & 1.8 & 1.3 & 1.3 & 1.7 &1.6 \\
   $1.0<M_p\leq10$     & 8.7 & 9.5 & 9.2 & 9.0 &8.4\\
   $10<M_p\leq100$     & 13.0 & 14.9 & 16.0 & 15.0 &13.8\\
   $100<M_p\leq1000$   & 11.3 & 13.9 & 15.0&  14.1 &12.8\\
   $1000<M_p\leq10000$ & 7.5  & 9.6 & 10.4 &  10.0 &9.0\\
  \hline 
  Total ($10^{-1}-10^4 M_\oplus$) &42.4 &49.1 & 51.8& 49.8 &45.6\\
  Total Microlensing         & $\sim 2300$ & $\sim3400$ & $\sim4100$ & $\sim3900$ & $\sim3400$\\
  \hline 
 \end{tabular}
  \end{threeparttable}

\end{table*}

{Table \ref{Table_planet_log} shows our estimation of the number of microlensing events and the number of planets detected by the PRIME microlensing survey assuming the \citet{2012Natur.481..167C} mass function as modified by \citet{2019ApJS..241....3P} (Equation \ref{eq-fp}) over a certain mass range. The total number of microlensing events detected are $\sim$ 2300, 3400, 4100, and 3900, for the S1, S2, S3, and S4 strategies, respectively. The impact of increasing the number of sources by observing more fields is more significant than the impact of decreasing the detection efficiencies by observing with {lower} cadence.}

{In Figure \ref{fig_Planet_yields}, we plot the planet detection rate per dex for four observation strategies, calculated by the sum of the semi-major axis over $0.3<a<30$ au and the sum of the survey area ($8.7 - 26.2$ deg$^2$) shown in Table \ref{Table_planet_log}).
In order to detect low mass planets, high cadence observations are required (S1), while in order to detect high mass planets, observing a larger number of fields is more important than observing with a higher cadence (S2 and S3). When we use a hybrid observation cadences (S4), it is possible to detect both low mass planets and high mass planets. The lower panel in Figure \ref{fig_Planet_yields} shows the detection rates of each strategy relative to that of S4.
As the result, we predict that PRIME will discover $42-52$ planets  ($1-2$ planets with $M_p \leq M_\oplus$,  $22-25$ planets with mass $1 M_\oplus < M_p \leq 100 M_\oplus$, $19-25$ planets  $100 M_\oplus < M_p \leq 10000 M_\oplus$), per year depending on each observation strategy.}

\clearpage
\section{Discussion}
\label{sec-dis}

\subsection{How to decide the optimal survey strategy?}\label{sec-dis-opt}

The final survey strategy will vary according to the interests of several sciences: to reveal the planet frequency around the Galactic center, to optimize the $Roman$ microlenisng survey fields, to characterize the lens and planet parameters by follow-up observations. We will discuss each of these science interests in detail.

In this paper, we focus on revealing the demography of cold planets down to Earth mass beyond the snow-line toward the inner Galactic bulge. In order to achieve that goal, it is required to optimize the observation strategy and {to increase both the number of planets and the range of mass comparing} four observation strategies, we find that it is possible to detect both low mass planets and high mass planets by an observation strategy with a hybrid observation cadence, S4. We predict that PRIME will discover up to $\sim 3900$ microlensing events and $\sim 50$ planets per year by using S4.

However another important goal of the PRIME is the optimization of the $Roman$ microlensing survey fields by measuring the NIR microlenisng event rate map and {$t_{\rm E}$ distributions}. In order to achieve that goal, it is required to conduct a spatially uniform survey toward the inner Galactic bulge. We investigate how the planet yields change with the uniform survey strategy. The bottom panel in Figure \ref{fig_AHmap} shows the considered field locations when we conduct a uniform survey including the Galactic center and the Galactic plane. Here, we use a hybrid observation cadence and the total number of fields is 18, which are the same as in observation strategy S4. Table \ref{Table_planet_log} shows our estimation of the number of microlensing events and the planet detections. The result shows $\sim6-10\%$ fewer planet discoveries depending on the planet mass and $\sim13\%$ fewer microlensing discoveries compared to the observation strategy, S4. Therefore, the uniform survey not only allows for the detection of a relatively large number of planetary signals including low-mass planets to measure the planet frequency toward the Galactic inner bulge, but also allows for the measurement of event rates across the Galactic center and Galactic plane to help optimize $Roman$'s observation strategy.

{NIR or optical follow-up observations will help to constrain the microlensing and physical parameters of planetary systems. In particular, color measurements of microlenisng events will enable us to determine $\theta_{\rm E}$, which constrains the lens mass and distance. Differences in extinction can affect field selection because they affect whether color measurements can be performed or not, but field selection by extinction in other bands is outside the scope this work.}

\subsection{Inner Galactic bulge survey by PRIME}

{In this study, we use KGM, which is a population synthesis model optimized for the inner Galactic bulge that includes a nuclear stellar disk model. 
As shown in Section \ref{sec-star-counts}, the luminosity function at the low mass stars is not in agreement with measurements.
It is also known that there is the underestimation of extinction values in the Galactic central region which is shown in Koshimoto et al. (in prep). }
{Observations of the star counts, event rate, and detection efficiencies will drive improvements in Galactic models.}

{Although previous NIR observations towards the inner Galactic bulge such as the VVV survey have revealed detailed structure of the Galactic bar/bulge (e.g. \citealp{2013MNRAS.435.1874W,2015MNRAS.450.4050W}), the formation history and structure of our Galaxy is a long-standing challenge \citep{2020RAA....20..159S}.
To constrain the dynamical history and evolution of Galaxy, accurate measurements of a stellar 6-D phase space distribution and stellar properties in the inner bulge region will be provided by the future time domain survey such as $Roman$, the {\it Japan Astrometry Satellite Mission for INfrared Exploration} ({\it JASMINE}; \citealp{2012ASPC..458..417G}) and {\it GaiaNIR} \citep{2016arXiv160907325H,2019arXiv190712535H}.
Prior to these surveys, a time domain survey with high cadence using PRIME will play an important role in providing new insights into the formation history and structure of our Galaxy. In addition to aspects of microlensing, the time domain data by the PRIME microlensing survey will provide useful information in studies of Galactic structure, through variable stars such as eclipsing binaries, pulsating RR Lyrae, and Cepheids (e.g. \citealp{2020AcA....70..121P,2021MNRAS.504..654B}).}

\section{Summary}
\label{sec-sum}
We present the expected microlensing and planet yields for four survey strategies using the PRIME instrument. In order to maximize the number of planet detections and the range of masses, we need to optimize the number of the observation fields and observation cadence, which are in a trade-off relationship. Assuming the an underlying planet population of one planet per square dex per star and the \citet{2012Natur.481..167C} mass function of planets beyond snow-line as modified by \citet{2019ApJS..241....3P}, we predict that PRIME will discover $2300-4100$ microlensing events and $42-52$ planets per year depending on the observation strategy. 
In particular, the observation strategy with a hybrid observation cadence (S4) makes it possible to detect both low mass planets and high mass planets. By using S4, we predict that PRIME will discover up to $\sim 3900$ microlensing events and $\sim 50$ planets per year ($\sim1.7$ planets with $M_p \leq 1 M_\oplus$,  $\sim24$ planets with mass $1 M_\oplus < M_p \leq 100 M_\oplus$, $\sim24$ planets  $100 M_\oplus < M_p \leq 10000 M_\oplus$). Besides, the spatially uniform survey not only allows for the detection of a relatively large number of planetary signals including low-mass planets, but also allows for the measurement of event rates across the Galactic center and Galactic plane.

\software{\texttt{genstars} (\citealp{kos22}; Koshimoto et al. in prep.),
          \texttt{MulensModel} \citep{2019A&C....26...35P},  
          \texttt{VBBinaryLensing} \citep{2010MNRAS.408.2188B, 2018MNRAS.479.5157B}
          }
          

We would appreciate Kento Masuda for valuable comments and discussions. 
Work by I.K. is supported by JSPS KAKENHI Grant Number 20J20633. 
Work by T.S. is supported by JSPS KAKENHI Grant Number 23103002, 24253004, and 26247023. 
Work by N.K. is supported by the JSPS overseas research fellowship.
Work by D.S. is supported by JSPS KAKENHI Grant Number 19KK0082 and 20H04754.
Work by D.P.B. is supported by NASA through grant NASA-80NSSC18K0274. 
This research has made use of the KMTNet system operated by the Korea Astronomy and Space Science Institute (KASI) and the data were obtained at three host sites of CTIO in Chile, SAAO in South Africa, and SSO in Australia.

\bibliography{reference}{}
\bibliographystyle{unsrt}
\bibliographystyle{aasjournal}

\end{document}